\def\today{\ifcase\month
\or January\or February\or March\or April\or May
\or June\or July\or August\or September\or October
\or November \or December\fi \space\number\day, \number\year}
\input feynman
\bigphotons
\def\bold#1{\setbox0=\hbox{$#1$}%
     \kern-.025em\copy0\kern-\wd0
     \kern.05em\copy0\kern-\wd0
     \kern-.025em\raise.0433em\box0 }
\def\slash#1{\setbox0=\hbox{$#1$}#1\hskip-\wd0\dimen0=5pt\advance
       \dimen0 by-\ht0\advance\dimen0 by\dp0\lower0.5\dimen0\hbox
         to\wd0{\hss\sl/\/\hss}}
\def\Gammait{{\mit\Gamma}}
 \def\Ybar{\overline{Y}}
 \def\Itil{\tilde{I}} \def\Btil{\tilde{B}}
\def\Stil{\tilde{S}}
\def\Gammat{\tilde{\Gamma}} \def\Omegat{\tilde{\Omega}} \def\Stil{\tilde{S}}

\def\msla{\slash m} \def\nsla{\slash n} \def\qsla{\slash q}
\def\ksla{\slash k} \def\psla{\slash p} \def\Vsla{\slash V}
\def\psib{\overline{\psi}} \def\Psib{\overline{\Psi}}
\def\Ccal{{\cal C}} \def\Lcal{{\cal L}} \def\Dcal{{\cal D}}
\def\Mcal{{\cal M}} \def\Acal{{\cal A}} \def\Jcal{{\cal J}}
\def\Fcal{{\cal F}} \def\Gcal{{\cal G}} \def\Acalt{\tilde{\Acal}}
\def\Qcal{{\cal Q}} \def\Rcal {{\cal R}} \def\Ecal{{\cal E}}
  
 \def\Ihat{{\hat{I}}} \def\Gammah{\hat{\Gamma}}
\def\gammah{\hat{\gamma}} \def\alphah{{\hat{\alpha}}}
\def\gammab{\overline{\gamma}}
\def\gammap{{1+\gamma_3\over2}} \def\gammam{{1-\gamma_3\over2}}
 
 \def\trace{\mathop{\rm tr}\nolimits}
\def\partialh{\hat{\partial}} \def\gs{\sqrt{g}} 
\def\eleft{{\grave e}} \def\gleft{{\grave g}} \def\gsleft{\sqrt{\gleft}}
\def\Aleft{{\grave A}} \def\Dleft{{\grave D}} 
\def\eright{{\acute e}} \def\gright{{\acute g}} \def\gsright{\sqrt{\gright}}
\def\Omegar{{\acute\Omega}} \def\Dright{{\acute D}}
\def\partialr{{\acute\partial}}
\newif\ifdraft
\documentstyle[12pt,equations,a4]{article}

\def\hcite#1{\cite{#1}\ifdraft\footnote{#1}\else\fi}
\def\tlabel#1{\label{#1}\ifdraft\mbox{\tiny \hsize4pt (#1)}\fi}
\def\hlabel#1{\label{#1}\ifdraft\ifmmode\mbox{\tiny \hsize4pt (#1)}
 \else\footnote{(#1)}\fi\else\fi}
\def\beb{\begin{subequations}}  \def\eeb{\end{subequations}}
\def\bgn{\begin{eqalign}}       \def\egn{\end{eqalign}}
\def\btw{\begin{eqaligntwo}}    \def\etw{\end{eqaligntwo}}
\def\bts{\begin{eqaligntwo*}}   \def\ets{\end{eqaligntwo*}}
\def\bar{\begin{eqnarray}}      \def\ear{\end{eqnarray}}
\def\bas{\begin{eqnarray*}}     \def\eas{\end{eqnarray*}}
\def\bno{\begin{eqalignno}}     \def\eno{\end{eqalignno}}
\def\bns{\begin{eqalignno*}}    \def\ens{\end{eqalignno*}}
\def\beq{\begin{equation}}      \def\eeq{\end{equation}}
\def\bes{\[}                    \def\ees{\]}

\begin{document}
\thispagestyle{empty}
\null
\vspace{-2truecm}
\rightline{\protect{{\small DFPD/94/TH/43}}}
\rightline{\protect{{\small hep-th/9409107}}}
\vspace*{.4truecm}
\begin{center}
\begin{large}\begin{bf}
k-ANOMALIES AND SPACE-TIME SUPERSYMMETRY\\
IN THE GREEN-SCHWARZ HETEROTIC SUPERSTRING\footnote{Supported
in part by M.P.I. This work is carried out in the framework of
the European Community Programme ``Gauge Theories, Applied Supersymmetry
and Quantum Gravity'' with a financial contribution under contract SC1-CT92
-D789.}
\\[1.5truecm]
\end{bf}\end{large}
A. Candiello, K. Lechner and M. Tonin\\[1truecm]
\begin{sl}
Dipartimento di Fisica, Universit\`a di Padova\\
and\\
Istituto Nazionale di Fisica Nucleare, Sezione di Padova\\
Italy\\[1truecm]
\end{sl}
\begin{bf}
Abstract
\end{bf}
\end{center}
The computation of $\kappa$-anomalies in the Green-Schwarz heterotic
superstring sigma-model and the corresponding Wess-Zumino consistency
condition constitute a powerful alternative approach for the derivation of
manifestly supersymmetric string effective actions. With respect to the
beta-function approach this technique presents the advantage that a
result which is obtained with the computation of beta-functions at $n$ loops
can be obtained through the calculation of $\kappa$-anomalies at \hbox{$n-1$}
loops. In this paper we derive by a direct one-loop perturbative computation
the $\kappa$-anomaly associated to the Yang-Mills Chern-Simons threeform and,
for the first time, the one associated to the Lorentz Chern-Simons threeform.
In the calculation we shall use a convenient set of constraints for
the pure $N=1$, $D=10$ supergravity theory which is algebraically identical to
the standard set of constraints for the pure $N=1$, $D=10$ super Yang-Mills
theory. Contrary to what is often stated in the literature we show that the
Lorentz $\kappa$-anomaly gets contributions from the integration over both the
fermionic {\it and\/} bosonic degrees of freedom of the string. A careful
analysis of the absolute coefficients of all these anomalies reveals that they
can be absorbed by setting $dH={\alpha'\over4}(\trace F^2-\trace R^2)$, where
$\alpha'$ is the string tension, the expected result. We show that this
relation ensures also the absence of gauge and Lorentz anomalies in the
sigma-model effective action. Moreover, the
consistency condition of the $\kappa$-anomalies ensures the closure of the
SUSY algebra in the Bianchi identities. We evidenciate the presence of
infrared divergences in the heterotic string sigma model, which are due to the
presence of the $d=2$ scalar massless fields of the string, and present a
conjecture for their cancellation which is intimately related to the locality
and Wess-Zumino consistency of the $\kappa$-anomalies.
\newpage
\baselineskip20pt
\setcounter{page}{1}
\baselineskip20pt
\setcounter{page}{1}
\section{Introduction and summary}
In its original formulation, the Neveu-Schwarz-Ramond (NSR) formulation,
superstring theory appears manifestly Lorentz-covariant in its critical $D=10$
dimension, while its principal drawback is the missing {\it manifest\/}
target space-time supersymmetry. Its alternative formulation, the
Green-Schwarz (GS) formulation, on the other hand exhibits manifest $D=10$
space-time supersymmetry but, despite a lot of efforts, no manifestly
Lorentz-covariant quantization scheme has been found until now. This
difficulty is due to the fact that $\kappa$-invariance, the fundamental
symmetry of the GS-string, cannot be fixed in a manifestly Lorentz-covariant
way.

In the low energy limit superstring theory reduces to an $N=1$, $D=10$
Supergravity-Super-Yang-Mills (SUGRA-SYM) theory whose dynamics is described
by appropriate effective actions. In the past such effective actions have been
derived directly from the string amplitudes (see e.g. \hcite{GROSS}) or by
imposing the vanishing of the beta functions in string sigma models embedded
in the zero modes of the string (see e.g. \hcite{GRIVEN,FRATSE}). These
methods have been
carried out  almost exclusively in the NSR formulation, and as such they miss
manifest space-time supersymmetry, but there is also some important work
carried out in the GS framework \hcite{GRIZAN,GRIZAN2,HAAG}.

Strings in the GS formulation on the other hand furnish an approach for the
derivation of manifestly supersymmetric effective actions which relies neither
on the knowledge of string amplitudes
nor on the computation of beta-functions but on the fundamental
$\kappa$-invariance of the GS-string. It goes as follows \hcite{TONIN}.
One writes a
string sigma model action embedded in the ten-dimensional SUGRA-SYM superspace
describing the massless modes of the underlying string theory. The
$\kappa$-invariance
of this action at the classical level implies constraints on the
background supercurvatures and torsions which via the Bianchi identities lead
to the equations of motion for the background fields; for the heterotic string,
under investigation in this paper, at zeroth order in the string coupling
constant $\alpha'$ these constraints describe the pure
minimal SUGRA decoupled from SYM and the flat $N=1$, $D=10$ SYM theories.
If one quantizes the sigma model, $\kappa$-anomalies can show up, whose form
is strongly restricted by the Wess-Zumino (WZ) consistency condition. It turns
out \hcite{TONIN} that the non
trivial solutions of the corresponding cohomology problem are all such that
the related $\kappa$-anomalies can be absorbed by suitably modifying the
(classical) constraints on the supercurvatures and torsions of the background
fields. This procedure has to be carried out order by order in $\alpha'$, i.e.
loop by loop in the quantum expansion of the sigma model.
The solution of the Bianchi identities with these new constraints
gives the new equations of motion and hence the string-corrected effective
action for SUGRA-SYM as a power series in $\alpha'$ with manifest SUSY.

In this paper we want to illustrate the powerfulness and conceptual elegance
of this procedure in the heterotic string by performing in particular for the
first time the direct perturbative computation of the $\kappa$-anomaly related
to the Lorentz-Chern-Simons term, but we shall also evidenciate its technical
problems
and conceptual limitations among which the most striking one is the appearance
of infrared divergences.

With respect to the beta-function approach our algorithm presents a decisive
advantage: a contribution to the effective action obtained with beta-functions
at the $n$-th loop order is obtained with our algorithm at $(n-1)$ loops. May
be this is the reason why the Lorentz-Chern-Simons term has never been derived
using beta-functions (a two-loop computation!) while the $\kappa$-anomaly
implying this term arises at one-loop. Another technical difficulty related
with the beta-function approach is that the absence of a manifestly Lorentz
covariant quantization procedure gives rise to Lorentz non-covariant
intermediate results which are not easy to handle. The $\kappa$-anomaly
algorithm, on the other hand, produces directly the modified constraints on the
superspace so that the equations of motion can be derived in a straightforward
way by solving the Bianchi identities via standard techniques, and the Lorentz
non-covariance of the quantization procedure can be easily handled, at least
in the computation of the Lorentz $\kappa$-anomaly.
\vskip.3truecm

The paper is organized as follows. In section II we present the pure
Supergravity and Super Yang-Mills system which constitutes the background of
the heterotic Green-Schwarz string, along with the constraints on the
supercurvatures and torsions needed to set the decoupled theory on-shell. The
set of
these constraints is not unique, but is determined modulo field redefinitions.
Using this freedom we choose for the SUGRA a set of constraints in which the
purely spinorial components of the Lorentz-curvature vanish,
$R_{\alpha\beta a}{}^b=0$ \hcite{CANLEC,BOPATO}. This is possible at the
classical level where SUGRA is decoupled from SYM. This constraint being
analogous to the constraint for the SYM curvature, $F_{\alpha\beta}=0$
\hcite{NILSSON}, when
computing the Lorentz $\kappa$-anomaly we can, to a certain extent but with
an important difference, follow the procedure for the computation of the
Yang-Mills $\kappa$-anomaly.

In section III we present the action for the Green-Schwarz heterotic sigma
model embedded in the SUGRA-SYM background together with its symmetries. The
action is $\kappa$-invariant only if the constraints that pose the background
fields on-shell are satisfied. The action is also invariant under gauge and
Lorentz combined transformations of the background fields and the string
fields. These transformations, as we will see, give rise to ``anomalies'', but
we would like to point out that these transformations, not being actually
symmetries of the theory, do not produce true anomalies: they are a useful
tool for the analysis of the related gauge- and Lorentz-type
$\kappa$-anomalies which are true anomalies of the theory.

In section IV we discuss briefly the quantization and describe the normal
coordinate expansion of the Green-Schwarz action in the framework of the
Batalin-Vilkovisky approach.

In section V we rederive the gauge anomaly and the $\kappa$-anomaly associated
to the Yang-Mills Chern-Simons form. Usually when computing an anomaly one
regularizes the classical action, computes the associated effective action and
gets the anomaly by varying the effective action. A less known alternative
method consists in regularizing the classical action and computing
the variation of the regularized classical action to get an ``anomalous
vertex''. The anomaly is simply obtained by inserting the anomalous vertex in
all Feynman diagrams and by keeping only those which survive when the
regulator goes to zero. We shall use this alternative method to compute the
gauge anomaly and the $\kappa$-anomaly associated to the Yang-Mills
Chern-Simons form.

In section VI we apply a $\kappa$-gauge fixing to the expanded action which
breaks the manifest Lorentz invariance of the theory, but leaves a residual
$SO(8)$ invariance there. In analogy to the computation of section V we
identify the anomalous vertex associated to Lorentz transformations
and compute the corresponding Lorentz anomaly together with its absolute
coefficient.  It turns out that only the fermionic degrees of freedom of the
string contribute to the Lorentz anomaly, and that its coefficient, with
respect to the naive guess, is divided by a factor of two. This is due to the
fact that the $\kappa$-symmetry implies that half of the 16 fermionic degrees
of freedom of the string are unphysical and therefore only 8 of them circulate
in the anomalous
diagram. The final result can be easily Lorentz-covariantized by employing the
manifest $SO(8)$ invariance of the result. The gauge and Lorentz anomalies
computed in sections V and VI can be eliminated by associating to the
two-superform potential $B$ of the $N=1$, $D=10$ supergravity sector
transformation properties such that its curvature, defined as
$H=dB+{\alpha'\over4}(\omega_{3YM}-\omega_{3L})$ where $\omega_{3YM}$ and
$\omega_{3L}$ are the Yang-Mills and Lorentz Chern-Simons forms, is gauge and
Lorentz invariant, as one expects.

Section VII is devoted to the computation of the Lorentz $\kappa$-anomaly. In
the Yang-Mills sector the $\kappa$-transformation acts essentially as a
field-dependent gauge transformation and therefore the computation of the
$\kappa$-anomaly is closely related to that of the gauge anomaly.
The action of the $\kappa$-transformation in the gravitational sector is,
however, a combination of a field-dependent local Lorentz
transformation and an ``intrinsic'' $\kappa$-transformation and the relation
between the Lorentz $\kappa$-anomaly $\Acal_\kappa$ and the Lorentz anomaly
$\Acal_L$ is less obvious. In fact ${\cal A}_L$ gets contributions
only from loops where the fermionic fields of the string circulate, while the
$\kappa$-anomaly ${\cal A}_\kappa$ gets contributions also from loops with the
bosonic fields of the string circulating. These loops are necessary to
saturate the coupled cohomology problem
\beq\bgn
\Omega_L{\cal A}_L&=0\\
\Omega_\kappa{\cal A}_L+\Omega_L{\cal A}_\kappa&=0\\
\Omega_\kappa{\cal A}_\kappa&=0.
\egn\hlabel{eq:coupled2}\eeq
where $\Omega_\kappa$ and $\Omega_L$ are the BRS operators associated to the
$\kappa$-transformations and Lorentz transformations respectively.
Here the situation is similar to that found in the case of the SUSY anomaly
${\cal A}_S$ in a supersymmetric chiral Yang-Mills theory. In that case the
presence of an ABBJ Yang-Mills anomaly ${\cal A}_G$ induces the presence of a
SUSY anomaly via the coupled cohomology problem:
\beq\bgn
\Omega_G{\cal A}_G&=0\\
\Omega_S{\cal A}_G+\Omega_G{\cal A}_S&=0\\
\Omega_S{\cal A}_S&=0,
\egn\hlabel{eq:coupled1}\eeq
where $\Omega_G$ and $\Omega_S$ are the BRS operators associated to gauge and
SUSY transformations respectively. As is well known the ABBJ anomaly
${\cal A}_G$ gets contributions only from one-loop diagrams where chiral
quarks circulate, while the SUSY anomaly ${\cal A}_S$ gets contributions also
from loops of squarks, the scalar bosonic superpartners of the quarks.
These loops with squarks are necessary to saturate the coupled
cohomology problem (\ref{eq:coupled1}).

We derive ${\cal A}_\kappa$ through the standard procedure by identifying the
relevant part of the effective action, by integrating over fermions {\it and\/}
bosons and by varying it. Besides the local terms which saturate exactly
(\ref{eq:coupled2}) one gets infrared divergences coming from the integration
over the {\it massless\/} bosons which are non local and which spoil,
moreover, $\kappa$-invariance in the sense that they would give rise to
non-local $\kappa$-anomalies. This is clearly related to the fact that there
exists no $\kappa$-symmetry preserving infrared regularization procedure for
the GS sigma model: as it stands, the perturbative expansion of the GS sigma
model effective action is inconsistent due to the presence of these infrared
divergences. We argue by exhibiting an explicit simplified example that these
divergences are actually due to an intrinsic non analyticity of the effective
action, as a functional of the fields, which can therefore not be expanded
perturbatively as a polynomial in the external fields. Assuming that a
non-perturbative treatment will eventually eliminate these divergences we can
invoke a) the Wess-Zumino consistency condition for the $\kappa$-anomalies and
b) their locality to eliminate them completely without arbitrariness left. But
this recipe amounts to a conjecture and not to a solution of the infrared
problem.

The $\kappa$-anomalies derived in this way induce a background SUGRA-SYM
theory based on the Bianchi identity in superspace
\beq
dH={\alpha'\over4}\left(\trace F^2-\trace R^2\right)
\hlabel{eq:3tex}\eeq
precisely as predicted by the Green-Schwarz anomaly cancellation mechanism in
$N=1$, $D=10$ SUGRA-SYM \hcite{GRESCH} and by the (non supersymmetric)
effective action derived directly by the Veneziano-like superstring amplitudes
\hcite{GROSS}.

The $\kappa$-anomaly method produces automatically the constraints on the
background fields with which one has to solve (\ref{eq:3tex}) once the WZ
consistency condition is satisfied. In section VIII we check that our
$\kappa$-anomalies satisfy indeed the WZ condition and determine the
corresponding superspace constraints. Differences between these constraints
and other constraints in the literature \hcite{RARIZA} are shown to be related
to $\kappa$-cocycles trivial at one loop.

Section IX contains some conclusions and outlooks on the $\kappa$-anomaly
computation at higher loop orders, together with a brief analysis of the open
problems in the quantization procedure and perturbative treatment of the GS
string sigma model.
\vspace{1truecm}
\section{Pure Supergravity and Super Yang-Mills}
In this section we outline the background theory required by the Green-Schwarz
heterotic sigma model.

A superspace in ten dimensions is parametrized by the coordinates
$Z^M(\sigma)=\left(X^m(\sigma),\vartheta^\mu(\sigma)\right)$, where $X^m$ ($m=
0,1,\ldots,9$) are the bosonic degrees of freedom and $\vartheta^\mu$ ($\mu=
1,\ldots,16$) are the fermionic degrees of freedom.
The supervielbein one-form $E^A=dZ^ME_M{}^A(Z)$ describes the local flat frame
($A=(a,\alpha)$, where ($a=0,1,\ldots,9$; $\alpha=1,\ldots,16$), is a flat
index). For the ten-dimensional local Lorentz group we use a Minkowski
metric $\eta_{ab}$ with signature $-8$.
The $SO(32)$ Lie-valued Yang-Mills connection one-superform is $A=E^BA_B(Z)$,
while the Lorentz-valued connection one-superform is
$\Omega_A{}^B=E^C\Omega_{CA}{}^B(Z)$, where
$\Omega_a{}^\alpha=\Omega_\alpha{}^a=0$,
$\Omega_\alpha{}^\beta={1\over4}(\Gamma^{ab})_\alpha{}^\beta\Omega_{ab}$. The
supergravity potentials also comprehend the two-superform
$B={1\over2}E^CE^DB_{DC}(Z)$.
The field strengths associated to $E^A$, $B$, $A$ and $\Omega_A{}^B$ are
given by
\beb\bno
T^A&=DE^A=dE^A+E^B\Omega_B{}^A\\
W&=dB\\
F&=dA+AA\\
R_A{}^B&=d\Omega_A{}^B+\Omega_A{}^C\Omega_C{}^B
\eno\eeb
and the corresponding Bianchi
identities are
\beb\bno
DT^A&=E^BR_B{}^A\\
DW&=0\\
\Dcal F&=0\\
DR_A{}^B&=0,
\eno\hlabel{eq:Bianchi}\eeb
where $d=dZ^M\partial_M$, $D$ is the Lorentz covariant superdifferential and
$\Dcal$ is the gauge covariant superdifferential.
The pure supergravity and Yang-Mills theories are set on-shell by imposing a
minimal set of constraints on the curvatures, which is uniquely determined
modulo field redefinitions, and we choose it to be
\beb\hlabel{eq:grconst}\bno
T_{\alpha\beta}{}^a&=2\Gamma^a_{\alpha\beta},\qquad T_{\alpha a}{}^b=0
\hlabel{eq:6a}\\
(dB)_{a\alpha\beta}&=2(\Gamma_a)_{\alpha\beta},\qquad
(dB)_{\alpha\beta\gamma}=0,\qquad(dB)_{\alpha ab}=0\hlabel{eq:6b}\\
F_{\alpha\beta}&=0\hlabel{eq:Fconst}\\
R_{\alpha\beta a}{}^b&=0.\hlabel{eq:Rconst}
\eno\eeb
Note in particular the constraint (\ref{eq:Rconst}): as shown in
\hcite{BOPATO,TONIN}, it can always be imposed for pure supergravity. This
constraint allows to maintain a close parallelism between the gauge and
gravitational sectors.

The Bianchi identities then imply \hcite{CANLEC}
\beb\hlabel{eq:7}\bno
T_{\alpha\beta}{}^\gamma&=2\delta_{(\alpha}^\gamma\lambda_{\beta)}
-(\Gamma_g)_{\alpha\beta}(\Gamma^g)^{\gamma\varepsilon}\lambda_\varepsilon\\
T_{a\alpha}{}^\beta&={1\over4}(\Gamma^{bc})_\alpha{}^\beta T_{abc}\\
W_{abc}&\equiv (dB)_{abc}=T_{abc}\\
D_\alpha\lambda_\beta&=-(\Gamma^g)_{\alpha\beta}D_g\phi
+\lambda_\alpha\lambda_\beta+{1\over12}(\Gamma^{abc})_{\alpha\beta}T_{abc}\\
D_\alpha T_{abc}&=-6(\Gamma_{[a})_{\alpha\varepsilon}T_{bc]}{}^\varepsilon\\
F_{a\alpha}&=2(\Gamma_a)_{\alpha\varepsilon}\chi^\varepsilon
\hlabel{eq:Fino}\\
R_{a\alpha bc}&=2(\Gamma_a)_{\alpha\varepsilon}T_{bc}{}^\varepsilon
\hlabel{eq:Rino}\\
D_\alpha\chi^\beta&={1\over4}(\Gamma^{ab})_\alpha{}^\beta F_{ab}
+T_{\alpha\varepsilon}{}^\beta\chi^\varepsilon\\
D_\alpha T_{cd}{}^\beta&={1\over4}(\Gamma^{ab})_\alpha{}^\beta R_{abcd}
+T_{\alpha\varepsilon}{}^\beta T_{cd}{}^\varepsilon.
\eno\eeb
Here $\chi^\varepsilon$ and $T_{ab}{}^\varepsilon$ are the gluino and the
gravitino field strengths, $T_{abc}$, the vectorial part of the torsion, is
completely antisymmetric in its indices, $\phi$ is the dilaton superfield and
the gravitello superfield is $\lambda_\alpha\equiv D_\alpha\phi$.
Note the symmetry between the gauge and Lorentz sector visible in the last
four equations. The computation of the related
equations of motion can now be performed (see for example, with constraints
slightly different from ours, Ref.~\hcite{RARIZA}), but for the purposes of
this work we do not need them.

It is also useful to introduce the gauge and Lorentz Chern-Simons
three-superforms
\beq\bgn
\omega_{3YM}&=\trace\left(AF-{1\over3}A^3\right)\\
\omega_{3L}&=\trace\left(\Omega R-{1\over3}\Omega^3\right)
\egn\hlabel{eq:csdef}\eeq
satisfying
\beq\bgn
d\omega_{3YM}&=\trace(FF)\\
d\omega_{3L}&=\trace(RR)
\egn
\hlabel{eq:csid}
\eeq
which will play a central role in what follows.
In (\ref{eq:csdef}), (\ref{eq:csid})
the traces are in the fundamental representations of $SO(32)$ and $SO(10)$
respectively.
\vspace{1truecm}
\section{The action and its symmetries}
The action for the heterotic Green-Schwarz sigma model in a SUGRA-SYM
background is given by \hcite{WITMEZ,ATICK}
\beq
I=-{1\over2}\int d^2\sigma\left(\gs g^{ij}V_i{}^aV_{ja}+
\varepsilon^{ij}V_i{}^CV_j{}^DB_{DC}-
\gs e_-{}^j\psi\Dcal_j\psi\right).
\hlabel{eq:action}\eeq
Our notations are as follows. The string worldsheet is parametrized by
the coordinates $\sigma^i$ ($i,j=0,1$). The sigma-model fields are the
zweibeins $e_\pm{}^i(\sigma)$ with $e_i{}^\pm$ its inverses,
the superspace coordinates $Z^M(\sigma)$ which are
worldsheet scalars and the 32 Majorana-Weyl heterotic world-sheet fermions
$\psi^r(\sigma)$ ($r=1,\ldots,32$) which stay in the fundamental
representation of $SO(32)$.
$\Dcal_j\psi=(\partial_j-A_j)\psi$, where $A_j=V_j{}^BA_B$ and the
induced supervielbein $V_i{}^A$ is defined as $V_i{}^A=\partial_iZ^ME_M{}^A$.
In the following we shall use flat light-cone indices on the worldsheet
defined by $W_\pm=e_\pm{}^iW_i$ if $W_i$ is a worldsheet vector.
The worldsheet metric is $g_{ij}(\sigma)$, with $g^{ij}$ its inverse and
$g=-\det g_{ij}$ and $\varepsilon^{ij}$ is the antisymmetric Ricci tensor. The
metric and the Ricci tensor can be expressed in terms of the zweibeins through:
\beq\bgn
g^{ij}&={1\over2}\left(e_-{}^ie_+{}^j+e_+{}^ie_-{}^j\right)\\
{\varepsilon^{ij}\over\gs}&={1\over2}\left(e_-{}^ie_+{}^j-e_+{}^ie_-{}^j\right).
\egn\eeq
The self-dual projector $P^{ij}=g^{ij}+\varepsilon^{ij}/\gs$ can be expressed
through the zweibeins as $P^{ij}=e_-{}^ie_+{}^j$. By introducing the
two-dimensional Dirac matrices $\gamma^p$ in a Majorana representation
$\gamma^0=\left(\begin{array}{cc}0&1\\1&0\end{array}\right)$,
$\gamma^1=\left(\begin{array}{cc}0&-1\\1&0\end{array}\right)$
such that $\gamma^3=-\gamma^0\gamma^1=
\left(\begin{array}{cc}-1&0\\0&1\end{array}\right)$ and
using two-component Majorana spinors $\psi$ to describe the heterotic fermions,
the last term in (\ref{eq:action}) can be written as
\beq
I_H={1\over2}\int d^2\sigma\gs e_-{}^j\psib\gamma_+\Dcal_j\psi
={1\over2}\int d^2\sigma\gs e_p{}^j\psib\gamma^p\gammap\Dcal_j\psi,
\hlabel{eq:12tex}\eeq
where $\gamma_\pm=\gamma^0\pm\gamma^1$ and $\psib=\psi^T\gamma^0$. For
notational simplicity we use the same symbol for one-component spinors since
no confusion should arise. The second term in (\ref{eq:12tex}) will be used in
the following.

The action (\ref{eq:action}) is invariant under $d=2$ diffeomorphisms
and local $d=2$ Lorentz and Weyl transformations. In addition the
Green-Schwarz action is also invariant under Siegel's local $\kappa$-symmetry
\hcite{SIEGEL} which permits to eliminate half of the 16 $\vartheta^\mu$. The
transformation parameter is a (self-dual) world-sheet vector and space-time
spinor $\kappa_{+\beta}(\sigma)$. The string fields transform as follows:
\beb\hlabel{eq:kappavara}\bno
\delta_\kappa Z^M&=\Delta^\alpha E_\alpha{}^M\\
\delta_\kappa\psi&=\Delta^\alpha A_\alpha\psi\equiv C\psi\\
\delta_\kappa e_+{}^i&=
-4e_-{}^i\left(V_+{}^\varepsilon
-{1\over2}\psi\chi^\varepsilon\psi\right)
\kappa_{+\varepsilon}\hlabel{eq:kgvar}\\
\delta_\kappa g&=\delta_\kappa e_-{}^i=0\hlabel{eq:kdgvar}
\eno\eeb
where
\beq
\Delta^\alpha=V_-{}^a(\Gamma_a)^{\alpha\beta}\kappa_{+\beta}
\equiv (\Vsla_-\kappa_+)^\alpha;\hlabel{eq:deltadef}
\eeq
we use the notation
${\slash W}\equiv W_a(\Gamma^a)_{\alpha\beta}$ for a vector field $W_a$.
Correspondingly it can be seen that the target superfields and superforms
transform as
\beb\hlabel{eq:kappavar}\bno
\delta_\kappa V_i{}^A&=D_i\Delta^\alpha\delta_\alpha{}^A+
V_i{}^B\Delta^\gamma T_{\gamma B}{}^A-V_i{}^BL_B{}^A\\
\delta_\kappa B_{MN}&=\delta_\kappa Z^L\partial_LB_{MN}\\
\delta_\kappa T_{A\cdots}{}^{B\cdots}&=
\Delta^\alpha \partial_\alpha T_{A\cdots}{}^{B\cdots}\\
\delta_\kappa A_i&={\cal D}_iC+F_i\hlabel{eq:kgaugevar}\\
\delta_\kappa\Omega_{ia}{}^b&=D_iL_a{}^b+R_{ia}{}^b
\eno\eeb
where we defined:
\beb\hlabel{eq:defs}\bno
\Omega_{ia}{}^b&=\partial_i Z^M\Omega_{Ma}{}^b\\
L_a{}^b&=\Delta^\gamma\Omega_{\gamma a}{}^b\\
F_i&=V_i{}^B\Delta^\alpha F_{\alpha B}\\
R_{ia}{}^b&=V_i{}^B\Delta^\alpha R_{\alpha Ba}{}^b.
\eno\eeb
Under $\kappa$-transformations the action varies as:
\bns
\delta_\kappa I&=-{1\over2}\int d^2\sigma\Big(
2\gs g^{ij}V_{ia}V_j{}^B\Delta^\gamma T_{\gamma B}{}^a+
\varepsilon^{ij}V_i{}^CV_j{}^D\Delta^\gamma (dB)_{\gamma DC}\\
&\qquad+\gs e_-{}^j\psi F_j\psi-4\gs V_-^2
\left(V_+{}^\varepsilon-{1\over2}\psi\chi^\varepsilon\psi\right)
\kappa_{+\varepsilon}\Big).
\yesnumber\hlabel{eq:actkvar}\ens
The vanishing of the purely gravitational
contribution in (\ref{eq:actkvar}) requires precisely the constraints
(\ref{eq:6a}), (\ref{eq:6b}).
With these constraints the gravitational part of (\ref{eq:actkvar}) becomes,
in fact,
\beq
(\delta_\kappa I)_{\rm grav}=2\int d^2\sigma
\left(\gs V_-^2V_+{}^\alpha\kappa_{+\alpha}
-\gs V_+{}^\alpha(\Vsla_-\Delta)_\alpha\right)
\eeq
which vanishes since from the definition (\ref{eq:deltadef}) of
$\Delta^\alpha$, one has
\beq
(\Vsla_-\Delta)_\alpha=V_-^2\kappa_{+\alpha},\qquad V_-^2\equiv V_-{}^aV_{-a}.
\hlabel{eq:delid}\eeq
The vanishing of the Yang-Mills
contribution in (\ref{eq:actkvar}) requires the vanishing of the
spinor-spinor component of the Yang-Mills curvature (\ref{eq:Fconst}).
Indeed with the aid of (\ref{eq:Fconst}) and (\ref{eq:Fino}) we get
\beq
F_i=-2V_i{}^b\Delta^\alpha(\Gamma_b)_{\alpha\beta}\chi^\beta
\eeq
and, due to (\ref{eq:delid}),
\beq
F_-=-2V_-^2(\kappa_{+\varepsilon}\chi^\varepsilon).
\hlabel{eq:Fid}\eeq

Notice that $\kappa$-invariance, at the classical level, does not imply any
particular constraint on the spinor-spinor components of the Lorentz
curvature two-superform $R_{\alpha\beta}$. There are, in fact, a lot of field
redefinitions which keep the constraints in (\ref{eq:6a}) and (\ref{eq:6b})
invariant and give rise to different choices for $R_{\alpha\beta a}{}^b$.
The constraint (\ref{eq:Rconst}) is extremely convenient for the purpose of
the computation of the Lorentz $\kappa$-anomaly. It allows to follow as
closely as possible the derivation of the Yang-Mills $\kappa$-anomaly. With
this respect we notice that the relations (\ref{eq:Rconst}) and (\ref{eq:Rino})
imply in complete analogy to the Yang-Mills case that
\beq
R_{iab}=-2V_i{}^c\Delta^\alpha(\Gamma_c)_{\alpha\beta}T_{ab}{}^\beta
\eeq
and therefore, due to (\ref{eq:delid})
\beq
R_{-ab}=-2V_-^2\kappa_{+\varepsilon}T_{ab}{}^\varepsilon
\hlabel{eq:Rid}\eeq
which is proportional to $V_-^2$, exactly as in (\ref{eq:Fid}).
Eq.~(\ref{eq:Rid}) will be of fundamental importance in the derivation of the
Lorentz $\kappa$-anomaly.
\vspace{1truecm}
\section{Quantization and normal coordinate expansion}
A preliminary step to quantize the sigma-model action considered in the
previous section is to gauge-fix its local symmetries. Since the algebra is
open and reducible  (in fact infinitely reducible) the safest way to do that
is to work in the Batalin-Vilkovisky (BV) approach \hcite{BATVIL}. Calling
$\phi^I$ all the fields, ghosts, antighosts, Lautrup-Nakanishi fields and
secondary ghosts of the model, one introduces for each $\phi^I$ an antifield
$\phi^*_I$ with statistics opposite to $\phi^I$ and writes the extended action
$S_0[\phi,\phi^*]$
\beq
S_0[\phi,\phi^*]=I[\phi]+(-1)^{n(I)}\phi^*_I\Delta^I[\phi,\phi^*].
\eeq
Here $n(I)$ is the grading of $\phi^I$.
$I[\phi]=S_0[\phi,0]$ is just the action in (\ref{eq:action}), and
the terms linear in $\phi^*_I$ are obtained by
coupling the antifields to the BRS transformations of the fields and the
higher order terms are chosen so that $S_0$ satisfies the master equation
\beq
(S_0,S_0)\equiv(-1)^{n(I)}
{\delta S_0\over\delta\phi^I}{\delta S_0\over\delta\phi^*_I}=0.
\hlabel{eq:30bism}\eeq
As usual, here and in the following,
repeated indices $I$ imply sums over
discrete indices and integration over worldsheet coordinates.
Notice that the BRS transformations of $\phi$ are
\beq
\delta\phi^I=(S_0,\phi^I)\big|_{\phi^*=0}=
(-1)^{n(I)}{\delta S_0\over\delta\phi^*_I}\Big|_{\phi^*=0}=\Delta^I[\phi,0].
\eeq
The formalism is a graded canonical one with $\phi$, $\phi^*$ as conjugate
variables and
\beq
(\Fcal,\Gcal)=(-1)^{n(I)}\left(
{\delta\Fcal\over\delta\phi^I}{\delta\Gcal\over\delta\phi^*_I}
+{\delta\Fcal\over\delta\phi^*_I}{\delta\Gcal\over\delta\phi^I}\right)
\eeq
as graded Poisson bracket, $\Fcal$ and $\Gcal$ being even functionals of
$\phi$, $\phi^*$ with zero ghost number. The gauge-fixing is realized through a
canonical transformation on $S_0[\phi,\phi^*]$, generated by a suitably
chosen ``gauge fermion'', $\Psi[\phi]$ of ghost number $-1$. We do not report
here the explicit
form of the extended action $S_0[\phi,\phi^*]$ for our heterotic string
sigma-model. It can be found for instance in the last paper of
Ref.~\hcite{TONIN}, Eq.~(3.6).

On the other hand, calculations of quantum effective actions are simplified
by using the background field technique. It consists in performing, before
doing the gauge fixing, a split of the field variables $\phi^I$ into a
classical part $\phi_0^I$ and their ``fluctuations'' $\chi^I$ to be
quantized. In order to maintain local Lorentz and gauge invariance we shall
adopt a variant of this method known as ``normal coordinate expansion''
\hcite{MUKHI,ATIDHA}. In that case the splitting is
\beq
\phi^I=\phi_0^I+\Phi^I(\phi_0,\chi)
\hlabel{eq:28f}\eeq
where $\chi^\Ihat$ are the quantum fields. More precisely let us divide the
set of fields $\phi^I$ in four groups
\beq
\phi^I\equiv\left(q^i(\sigma),\psi^r(\sigma),Z^M(\sigma),
k^\alphah_n(\sigma)\right)
\eeq
where $Z^M$ are the string supercoordinates, $\psi^r$ the heterotic fermions,
$q^i$ denote fields, ghosts etc that are inert under Lorentz and gauge
transformations and $k^\alphah_n$ are ghosts and LN fields that transform as
(left-handed or right-handed) Lorentz spinors (i.e. $\alphah$ denotes an upper
or lower index $\alpha$).

Similarly
\beq
\phi_0^I\equiv\left(q_0^i(\sigma),\psi_0^r(\sigma),Z_0^M(\sigma),
k^\alphah_{0n}(\sigma)\right)
\eeq
and
\beq
\chi^\Ihat\equiv\left(Q^i(\sigma),\Psi^r(\sigma),y^A(\sigma),
\kappa^\alphah_n(\sigma)\right).
\eeq
Then Eq.~(\ref{eq:28f}) writes
\beb\hlabel{eq:32f}\bno
q^i&=q_0^i+Q^i\hlabel{eq:32af}\\
Z^M&=Z_0^M+\Pi^M(Z_0,y)\\
\psi&=e^{\Lambda(Z_0,y)}\left(\psi_0+\Psi\right)\hlabel{eq:32cf}\\
k_n&=e^{\Sigma(Z_0,y)}\left(k_{0n}+\kappa_n\right)
\eno\eeb
where $\Pi^M$, $\Lambda$ and $\Sigma$ depend on $Z_0^M$ and $y^A$ only,
$\Lambda$ being $SO(32)$ Lie algebra valued and $\Sigma$ Lorentz valued. In
particular for the zweibein we write (\ref{eq:32af}) as
\beq
e_\pm{}^i=e_{0\pm}{}^i+h_\pm{}^i.
\eeq
It is possible to implement the normal coordinate expansion in the framework
of the BV approach, as will be seen elsewhere \hcite{PREP}. For our purposes it
is sufficient to sketch the procedure.

First notice that, after the splitting (\ref{eq:28f}), the action acquires an
invariance under a local shift of the background fields $\phi_0$, supplemented
by a suitable transformation of the quantum fields $\chi$. Then consider the
action
\beq
\Stil_0=S_0[\phi,\phi^*]+(-1)^{n(I)}\phi_{0I}^*\Ecal^I
\eeq
where $\phi_{0I}^*$ are the antifields for $\phi_0^I$ and $\Ecal^I$ are
the (classical) local shift ghosts.

The next step is to perform on $\Stil_0$ a canonical transformation of the
fields $\phi$, $\phi_0$ and their (conjugate) antifields $\phi^*$, $\phi^*_0$
to implement the transformation (\ref{eq:28f}) on the fields $\phi^I$, leaving
unchanged the background fields $\phi_0^I$. Then the gauge fixing is performed
by means of a further canonical transformation generated by a suitable gauge
fermion to obtain the final extended classical action
$S[\chi,\chi^*;\phi_0,\phi_0^*]$ where $\chi^*_\Ihat$ are the antifields
associated to $\chi^\Ihat$.

Path-integrating over $\chi^\Ihat$, one can define, by the standard procedure,
the effective action $\Gammat[\chi,\chi^*;\phi_0,\phi_0^*]$ (as usual,
the classical fields associated to the quantum fields $\chi^\Ihat$ are still
denoted $\chi^\Ihat$).

Thanks to the shift symmetry, it is possible to perform on $\Gammat$ a
canonical transformation to get an action $\Gammah[\chi,\chi^*;\phi_0,\phi^*]$
where the terms linear in $\chi^\Ihat$ are absent. Then by taking $\Gammah$ at
$\chi=0=\chi^*$ and $\Ecal=0$ one arrives at an effective action
$\Gamma[\phi_0,\phi_0^*]$ that satisfies the Slavnov-Taylor identity
\beq
(\Gammait,\Gammait)=0.
\hlabel{eq:37m}\eeq
The field equations are
\beq
{\delta\Gammait\over\delta\phi_0^I}[\phi_0,\phi^*_0]=0.
\eeq
At zeroth order in $\alpha'$, for $\phi^*_0=0$ and disregarding the ghost
fields one has the classical field equations
\beb\hlabel{eq:motion}\bno
D_-V_{+a}
+\psib_0\gamma_+\left(V_-{}^\alpha(\Gamma_a)_{\alpha\beta}\chi^\beta
+{1\over2}V_-{}^cF_{ac}\right)\psi_0&=0\\
\Vsla_{-\alpha\beta}
\left(V_+{}^\beta-{1\over2}\psi_0\chi^\beta\psi_0\right)&=0\\
\gamma_+\left(\partial_--A_-+
{1\over2\gs}\partial_i(\gs e_-{}^i)\right)\psi_0&=0\\
V_-^2&=0\hlabel{eq:virasoro}\\
V_+{}^aV_{+a}-\psi_0\Dcal_+\psi_0&=0.
\eno\eeb

We will limit ourselves to perform one-loop computations for an on-shell
configuration of the background fields $\phi_0$ satisfying (\ref{eq:motion}).
Notice that in particular, due to the
Virasoro constraint (\ref{eq:virasoro}), the vectors $F_i$, $R_{ia}{}^b$,
appearing in the transformation of the connections, become chiral
\beq\bgn
g_0^{ij}F_j&={\varepsilon^{ij}\over\sqrt{g_0}}F_j,
\qquad\mbox{{\rm i.e.}}\qquad F_-=0\\
g_0^{ij}R_{ja}{}^b&={\varepsilon^{ij}\over\sqrt{g_0}}R_{ja}{}^b,
\qquad\mbox{{\rm i.e.}}\qquad R_{-a}{}^b=0.
\egn\hlabel{eq:FRdual}\eeq
The normal coordinate expansion amounts to a suitable choice of the functions
$\Pi^M(Z_0,y)$, $\Lambda(Z_0,y)$ and $\Sigma(Z_0,y)$ in
Eqs.~(\ref{eq:32f}) in such a way as to
restore the Lorentz and gauge covariance of the expansion along the quantum
fields of a functional like the action $I$, Eq.~(\ref{eq:action}). The
geometrical meaning of $\Pi^M$ is that it defines the variables $y^A$ so that
$y^A$ are tangent vectors to the geodesic joining the origin of the normal
coordinate $Z_0$ to the point $Z$. For more details about $\Pi^M$ and $\Lambda$
see Ref.~\hcite{MUKHI} and \hcite{KASCTS} respectively. Up to second order
in $y^A$
\beb\bno
\Pi^M&=y^BE_B{}^M+{1\over2}y^By^CD_CE_B{}^M+o(y^3)\\
\Lambda&=y^BA_B+{1\over2}y^By^CD_CA_B+o(y^3),\hlabel{eq:55old}
\eno\eeb
where $D_C$ is the Lorentz covariant derivative. A scalar functional which, as
the action (\ref{eq:action}), depends on $Z^M$ only through $V_i{}^A(Z)$ and
the flat components of the connections and curvatures can now be expanded,
according to the Mukhi algorithm,
\beq
I(Z,\psi,q)=\sum_{n=0}^\infty{1\over n!}\Delta^nI(Z_0,\psi_0+\Psi,q_0+Q)
\eeq
where the repeated application of the operator $\Delta$ is defined as follows
\beb\bno
\Delta V_i{}^A&=D_iy^A+V_i{}^By^CT_{CB}{}^A\\
\Delta\Omega_{iA}{}^B&=V_i{}^Cy^DR_{DCA}{}^B\\
\Delta A_i&=V_i{}^Cy^DF_{DC}\\
\Delta T_{A\cdots}{}^{B\cdots}&=y^C\Dcal_CT_{A\cdots}{}^{B\cdots}\\
\Delta y^A&=0\\
\Delta(\psi_0+\Psi)&=0\\
\Delta(q_0+Q)&=0.
\eno\eeb
Here $D_iy^A=\partial_iy^A+y^B\Omega_{iB}{}^A$ and $T_{A\cdots}{}^{B\cdots}$
is any Lorentz and Yang-Mills tensor. The expansion of $V_i{}^A(Z)$ up to
second order in $y^A$ is
\bns
V_i{}^A(Z)&=\partial_iZ^ME_M{}^A(Z)=\partial_iZ_0^ME_M{}^A+D_iy^A
+\partial_iZ_0^ME_M{}^Cy^BT_{BC}{}^A\\
&+{1\over2}D_iy^Cy^BT_{BC}{}^A+
{1\over2}\partial_iZ_0^ME_M{}^Dy^ET_{ED}{}^Cy^BT_{BC}{}^A\\
&+{1\over2}\partial_iZ_0^ME_M{}^Cy^By^DD_DT_{BC}{}^A
+{1\over2}y^D\partial_iZ^M_0E_M{}^Cy^BR_{BCD}{}^A+o(y^3).\yesnumber
\hlabel{eq:34m}\ens
The fields on the last member of this expression are all evaluated in $Z_0$.
The action is still BRS invariant after normal coordinate expansion if we
maintain for the background fields $\phi_0$ the classical variations
(\ref{eq:kappavara}), (\ref{eq:kappavar}) and impose suitable transformation
properties on the quantum fields. These latter can be read from the terms
linear in $Q^*$, $\Psi^*$, $y_A^*$, in the action
$S[\chi,\chi^*;\phi_0,\phi^*_0]$. However a convenient way to get
$\delta_\kappa y^A$, $\delta_\kappa\Psi$, $\delta_\kappa h_p{}^i$ is the
following. Consider the expansion of $\delta Z^ME_M{}^A(Z)$, obtained in
analogy with the expansion of $V_i{}^A(Z)$ in Eq.~(\ref{eq:34m}) to obtain
\bns
\delta Z^ME_M{}^A(Z)&=\delta Z_0^ME_M{}^A
+(\delta y^A+y^D\delta Z_0^ME_M{}^E\Omega_{ED}{}^A)
+\delta Z_0^ME_M{}^Cy^BT_{BC}{}^A\\
&+{1\over2}(\delta y^C+y^D\delta Z_0^ME_M{}^E\Omega_{ED}{}^C)y^BT_{BC}{}^A+
{1\over2}\delta Z_0^ME_M{}^Dy^ET_{ED}{}^Cy^BT_{BC}{}^A\\
&+{1\over2}\delta Z_0^ME_M{}^Cy^By^DD_DT_{BC}{}^A
+{1\over2}y^D\delta
Z^M_0E_M{}^Cy^BR_{BCD}{}^A+o(y^3).\yesnumber\hlabel{eq:kexp}
\ens
Once the left-hand side of this equation is known and once one specifies
$\delta Z_0^M$ this equation can be perturbatively solved for $\delta y^A$.
For $\kappa$-transformations we have
\beb\hlabel{eq:ktwo}\bno
\delta_\kappa Z_0^ME_M{}^A(Z_0)&=\Delta^A(Z_0)\hlabel{eq:ktwo1}\\
\delta_\kappa Z^ME_M{}^A(Z)&=\Delta^A(Z)=\Delta^A(Z_0)+
y^BD_B\Delta^A(Z_0)+o(y^2)
\eno\eeb
where $\Delta^a=0$ and $\Delta^\alpha$ is given in Eq.~(\ref{eq:deltadef}).
Notice that $\delta y^A$ appears in (\ref{eq:kexp}) always in the combination
$\delta y^A+y^B\Delta^\gamma\Omega_{\gamma B}{}^A$
and that all other terms are Lorentz-covariant. Therefore we can solve this
equation perturbatively to get a Lorentz-covariant expression for this
combination. With the aid of (\ref{eq:ktwo}) we obtain the
$\kappa$-transformations for $y^A$ which, together with (\ref{eq:ktwo1}),
leave the expanded action invariant:
\beb\hlabel{eq:kyvar}\bno
\delta_\kappa y^a&=-y^c\Delta^\gamma\Omega_{\gamma c}{}^a-
\Delta^\gamma y^BT_{B\gamma}{}^a+o(y^2)\\
\delta_\kappa y^\alpha&=-y^\beta\Delta^\gamma\Omega_{\gamma\beta}{}^\alpha
-\Delta^\gamma y^BT_{B\gamma}{}^\alpha+y^B\Dcal_B\Delta^\alpha+o(y^2).
\eno\eeb
The $o(y^2)$ terms are all Lorentz-covariant. So we see that on the $y$'s
$\kappa$-transformations can be considered as a combination of a
field-dependent Lorentz-transformation, with parameter
$L_a{}^b\equiv\Delta^\gamma\Omega_{\gamma a}{}^b$, and an ``intrinsic''
Lorentz-preserving $\kappa$-transformation.

The BRS transformations on $\Psi$ can be obtained in a similar way. We
write (see (\ref{eq:32cf})):
\bns
\big(\partial_i-&A_i(Z)\big)\psi=e^\Lambda\Big[
(\partial_i-A_i(Z_0))(\psi_0+\Psi)-\Big(V_i{}^Ay^BF_{BA}+\\
&+{1\over2}\left(D_iy^A+V_i{}^Cy^DT_{DC}{}^A\right)y^BF_{BA}
+{1\over2}V_i{}^Ay^By^C\Dcal_CF_{BA}+o(y^3)\Big)(\psi_0+\Psi)\Big].\yesnumber
\ens
For generic variations $\delta\psi$, $\delta\psi_0$, $\delta\Psi$,
$\delta Z^M$, $\delta Z_0^M$, $\delta y^A$ we get therefore
\bns
\delta\psi-&\delta Z^MA_M\psi=e^\Lambda\Big[
\delta\psi_0-\delta Z_0^MA_M\psi_0+\delta\Psi
-\delta Z_0^MA_M\Psi-\\
&-\Big(\delta Z_0^ME_M{}^Ay^BF_{BA}
+{1\over2}(\delta y^A+y^E\delta Z_0^M\Omega_{ME}{}^A
+\delta Z_0^ME_M{}^Cy^DT_{DC}{}^A)y^BF_{BA}\\
&+{1\over2}\delta Z_0^ME_M{}^Ay^By^C\Dcal_CF_{BA}+
o(y^3)\Big)(\psi_0+\Psi)\Big].\yesnumber\hlabel{eq:pkfull}
\ens
If we apply this formula to $\kappa$-transformations we see that the l.h.s.
vanishes identically. On $\psi_0$ we impose its classical
$\kappa$-transformation
\beq
\delta_\kappa\psi_0=C\psi_0,
\eeq
$\delta_\kappa Z^M_0$ is known and $\delta_\kappa y^A$ has been determined
above. Notice that again only the Lorentz-covariant combination
$\delta_\kappa y^A+y^B\Delta^\gamma\Omega_{\gamma B}{}^A$ appears. Therefore
(\ref{eq:pkfull}) determines the $\kappa$-transformation of the quantum
heterotic fermions:
\bns
\delta_\kappa\Psi&=C\Psi+\big(\Delta^\alpha y^bF_{b\alpha}
+{1\over2}y^CD_C\Delta^\alpha y^bF_{b\alpha}\\
&+{1\over2}\Delta^\alpha y^by^C\Dcal_CF_{b\alpha}+o(y^3)\big)
(\psi_0+\Psi).\yesnumber\hlabel{eq:kpvar}
\ens
Also in this case we see that on the quantum fields $\Psi$ the
$\kappa$-transformations act as a field-dependent gauge transformation, with
transformation parameter $C=\Delta^\alpha A_\alpha$, plus an ``intrinsic''
gauge and Lorentz covariant $\kappa$-transformation.

The BRS transformation of the quantum zweibeins $h_\pm{}^i$ can be
obtained by expanding (\ref{eq:kgvar}) and (\ref{eq:kdgvar}) and demanding
again that $e_{0\pm}{}^i$ transforms ``classically''. The
$\kappa$-transformations are given by
\beb\hlabel{eq:khvar}\bno
\delta_\kappa h_-{}^i&=0\\
\delta_\kappa h_+{}^i&=-4e_{0-}{}^i
\left(D_+y^\varepsilon+V_+{}^By^CT_{CB}{}^\varepsilon
+{1\over2}\psi_0y^A\Dcal_A\chi^\varepsilon\psi_0\right)
\kappa_{+\varepsilon}-\nonumber\\
&\qquad-4h_-{}^i
\left(V_+{}^\varepsilon-{1\over2}\psi_0\chi^\varepsilon\psi_0\right)
\kappa_{+\varepsilon}+o(y^2).
\eno\eeb
Now we have to be more specific about our gauge-fixing choice.

To fix world-sheet diffeomorphisms, Weyl and Lorentz invariance we shall
impose the condition
\beq
h_\pm{}^i=0
\eeq
on the zweibeins quantum fields.

For what concerns $\kappa$-invariance, until now no $D=10$ Lorentz-preserving
quantization procedure is known. Therefore, as unpleasant as it may be, we are
obliged to resort to a non-covariant gauge-fixing \hcite{KALLOSH}.
Consequently we shall fix
$\kappa$-symmetry by introducing two light-like ten-dimensional constant
vectors $m^a$, $n^a$ satisfying
\beq\bgn
m^an_a&={1\over2}\\
m^am_a&=0=n^an_a
\egn\eeq
such that the matrices $\slash m\equiv m_a\Gamma^a$, $\slash n\equiv n_a
\Gamma^a$ can be used to project $SO(10)$ spinors down to $SO(8)$ spinors.
We impose
\beq
\slash n_{\alpha\beta}y^\beta=0
\hlabel{eq:gfix}\eeq
and restrict the background-connection $\Omega_{ia}{}^b(Z_0)$ according to
\beq
\Omega_{iab}n^b=0=\Omega_{iab}m^b
\hlabel{eq:ofix}\eeq
such that the covariant derivative preserves (\ref{eq:gfix})
\beq
\slash n_{\alpha\beta}D_iy^\beta=0.
\eeq
As a consequence of (\ref{eq:ofix}) we will get an $SO(8)$-invariant effective
action and can finally use this residual $SO(8)$ invariance to covariantize
our results back to $SO(10)$. This procedure supposes that in principle an
$SO(10)$ Lorentz-covariant quantization scheme is available.

As for the huge series of secondary symmetries which arise due to the
(infinite) reducibility of $\kappa$-symmetry, they will be fixed by imposing
on the quantum fields of the $\kappa$-ghosts, antighosts, LN fields and
secondary ghosts conditions like Eq.~(\ref{eq:gfix}) involving alternatively
the constant vectors $m^a$ and $n^a$ \hcite{KALLOSH}. These conditions
together with the relevant field equations imply that the whole chain of
$\kappa$-ghosts do not propagate in our gauge and can be disregarded at the
quantum level.

However the ghosts and antighosts of diffeomorphisms do propagate and in a
complete treatment they should be taken into account carefully. Yet in this
paper we are interested only on the $\kappa$-anomaly (at one loop) and the
diffeomorphisms ghosts are expected not to contribute with this respect.

We end this section by giving the normal-coordinate-expanded lagrangian at
second order in the quantum fields which is needed for our one-loop
computations. In performing the expansion along the
quantum variables we need also the relations (\ref{eq:7}) stemming from the
solution of the Bianchi identities with the constraints (\ref{eq:grconst}). We
get (for $h_\pm^i=0$)
\bns
L_2&=\gs\Big[y^\alpha V_-{}^a(\Gamma_a)_{\alpha\beta}D_+y^\beta
-{1\over2}D_-y_aD_+y^a
-2D_-y^aV_+{}^\beta y^\alpha(\Gamma_a)_{\alpha\beta}\\
&+2V_-{}^aV_+{}^by^\alpha T_{cb}{}^\varepsilon
(\Gamma_a)_{\varepsilon\alpha}y^c
-{1\over4}V_-{}^aV_+{}^\alpha y_dy^\gamma(\Gamma_a)_{\alpha\varphi}
(\Gamma_{bc})^\varphi{}_\gamma T^{bcd}\\
&-{1\over2}V_-{}^aV_+{}^by^dy^cR_{dacb}+{1\over2}D_-y^aV_+{}^bT_{gba}y^g
-V_-{}^\beta V_+{}^bT_{cb}{}^\alpha y^cy^g(\Gamma_g)_{\alpha\beta}\\
&+{1\over4}V_-^aV_+^by^\delta y^\gamma(\Gamma_{acd})_{\delta\gamma}T^{cd}{}_b
+V_-^aV_+{}^\alpha T_{\delta\alpha}{}^\beta
(\Gamma_a)_{\beta\gamma}y^\delta y^\gamma-2V_-{}^\alpha V_+{}^\beta
(\Gamma^g)_{\alpha\delta}(\Gamma_g)_{\beta\gamma}y^\gamma y^\delta\\
&+{1\over2}\Psi\Dcal_-\Psi\Big]+o(\Psi y)+o(\psi_0^2y^2).\yesnumber
\hlabel{eq:fullexp}\ens
The $o(\Psi y)+o(\psi_0^2y^2)$ terms will not enter our calculations so we did
not write them explicitly.

In the next section we will start doing one-loop computations.
\vspace{1truecm}
\section{A non standard derivation of the Yang-Mills anomaly and the related
$\kappa$-anomaly}
The normal-coordinate expanded lagrangian is also invariant under Yang-Mills
and Lorentz gauge transformations involving both the background fields and the
quantum fields. The Yang-Mills transformations are
\beq\bgn
\delta_G A_i&=\Dcal_i\Ccal\equiv\partial_i\Ccal+\Ccal A_i-A_i\Ccal\\
\delta_G\Psi&=\gammap\Ccal\Psi,
\egn\hlabel{eq:gaugevar}\eeq
where we have reintroduced the two-component notation for the quantum
heterotic fermions, and the local Lorentz transformations are
\beq\bgn
\delta_L\Omega_{iA}{}^B&=D_i\Lcal_A{}^B\equiv\partial_i\Lcal_A{}^B+
\Lcal_A{}^C\Omega_{iC}{}^B-\Omega_{iA}{}^C\Lcal_C{}^B\\
\delta_L y^A&=-y^B\Lcal_B{}^A\\
\delta_L V_i{}^A&=-V_i{}^B\Lcal_B{}^A\\
\delta_L T_{A\cdots}{}^{B\cdots}&=\Lcal_A{}^CT_{C\cdots}{}^{B\cdots}
-T_{A\cdots}{}^{C\cdots}\Lcal_C{}^B+\cdots
\egn\eeq
where $\Ccal$ is a local Lie algebra-valued parameter,
$\Ccal=\Ccal^IT^I$ and $\Lcal$ is a Lorentz-valued parameter,
$\Lcal_a{}^\alpha=\Lcal_\alpha{}^a=0$,
$\Lcal_\alpha{}^\beta={1\over4}(\Gamma^{ab})_\alpha{}^\beta\Lcal_{ab}$;
$\Dcal_i$ and $D_i$ are the gauge and Lorentz induced covariant derivatives,
respectively and $T_{A\cdots}{}^{B\cdots}$ is any Lorentz tensor.

As a consequence it is meaningful to speak of the anomalies of these
symmetries. The consideration of these Yang-Mills and Lorentz anomalies is a
useful tool to discuss the $\kappa$-anomalies associated to Yang-Mills and
Lorentz Chern-Simons forms, on which we are interested in this paper.

In this section we want to compute the by now well understood gauge-anomaly
of the Green-Schwarz sigma model in dimensional regularization by a
non standard method \hcite{TONAOY}. This rederivation of the gauge anomaly
will clarify also some aspect of the appearance of the $\kappa$-anomaly
associated to the Yang-Mills Chern-Simons form and guide us also in the
derivation of the Lorentz anomaly and the $\kappa$-anomaly associated to the
Lorentz Chern-Simons form.

Our computational method is based on the following rather general
consideration. Consider an action $I[\chi,\phi_0]$ which depends on a set of
external fields $\phi_0$, and on a set of quantum fields $\chi$ over which
we are going to perform a path integration. Let us moreover assume that the
action is at the classical level invariant under a set of transformations
$\delta\phi_0$, $\delta\chi$ with associated BRS charge $\Omega$
\beq
\Omega I=0.
\hlabel{eq:50m}\eeq
$\Omega$ is a nilpotent operator if the algebra of the symmetry
transformations is closed, but when the algebra closes only on-shell (open
algebra), as is the case of $\kappa$-transformations, $\Omega$ is nilpotent
only on-shell. In the Batalin-Vilkovisky approach Eq.~(\ref{eq:50m}) is
replaced by the master equation (\ref{eq:30bism}) for the extended action.

The Slavnov operator $(S,\cdot)$ is nilpotent in all cases once $S$ satisfies
the master equation. When the action $I$ (the extended action $S$) is
regularized dimensionally, going in $d=2+\epsilon$ dimensions, one gets an
action $I_\epsilon$ ($S_\epsilon$) which is no longer invariant (no longer
satisfies the master equations) if the regularization breaks the symmetry
\beq
(S_\epsilon,S_\epsilon)=\Qcal_\epsilon=\epsilon \Rcal_\epsilon
\hlabel{eq:51m}\eeq
or
\beq
\Omega I_\epsilon=Q_\epsilon=\epsilon R_\epsilon
\eeq
where $R_\epsilon=\Rcal_\epsilon\big|_{\chi^*=\phi_0^*=0}$. If $\epsilon\to0$
$I_\epsilon\to I$ ($S_\epsilon\to S$) and $Q_\epsilon\to0$
($\Qcal_\epsilon\to0$).

It is convenient to define an action $S_\epsilon^\eta$, introducing an
anticommuting constant parameter $\eta$ which at the end will be set to zero,
according to
\beq
S_\epsilon^\eta=S_\epsilon+\eta\Rcal_\epsilon
\eeq
and Eq.~(\ref{eq:51m}) becomes
\beq
(S_\epsilon^\eta,S_\epsilon^\eta)=
\epsilon{\delta S_\epsilon^\eta\over\delta\eta}.
\eeq
The effective action $\Gammait_\epsilon^\eta$ no longer satisfies the
Slavnov-Taylor identity (\ref{eq:37m}) which is now replaced by
\beq
(\Gammait_\epsilon^\eta,\Gammait_\epsilon^\eta)=
\epsilon{\delta\Gammait_\epsilon^\eta\over\delta\eta}
\eeq
or
\beq
\Omega\Gamma_\epsilon^\eta=
\epsilon{\delta\Gamma_\epsilon^\eta\over\delta\eta}
\hlabel{eq:prec}\eeq
where $\Gamma_\epsilon^\eta=\Gammait_\epsilon^\eta\big|_{\phi_0^*=0}$.

Due to the analyticity of the dimensional regularization at first order in
$\alpha$ ($\alpha=2\pi\alpha'$ plays here the role of Planck's constant)
i.e. at one loop, we can make the following expansion in $\eta$
\beq
\Gamma^\eta_\epsilon=I_\epsilon^\eta+\alpha\left(
\left(\Gamma_1+{1\over\epsilon}\Gamma_0\right)+
\eta\left(\Delta_1+{1\over\epsilon}\Delta_0\right)\right)
\hlabel{eq:succ}
\eeq
where $\Gamma_1$, $\Delta_1$ are finite and $\Gamma_0$, $\Delta_0$ parametrize
the divergent local contributions to the effective action. Putting this into
(\ref{eq:prec}) and setting then $\eta=0$ we get for the regularized physical
effective action
\beq
\Omega\Gamma_\epsilon=\epsilon\left( R_\epsilon+\alpha(\Delta_1+
{1\over\epsilon}\Delta_0)\right),
\eeq
and for $\epsilon\to0$
\beq
\Omega\Gamma=\alpha\Delta_0.
\hlabel{eq:40}
\eeq
The Wess-Zumino consistency condition is then
\beq
\Omega\Delta_0=0.
\hlabel{eq:wzcons}\eeq
If $\Delta_0$ cannot be written as the $\Omega$-variation of a local action it
constitutes an anomaly.

Here we note that, thanks to (\ref{eq:40}) and (\ref{eq:succ})
1) The anomaly $\Delta_0$ is local and finite;
2) the divergent part of the effective action is BRS invariant.
What we learned from these considerations, taking a
look at (\ref{eq:succ}), is that the anomaly can be computed by inserting the
``anomalous vertex'' $ R_\epsilon$ once in all one-loop diagrams and keeping
the
$1/\epsilon$-divergent contributions or, alternatively, by inserting
$Q_\epsilon$ and taking the limit for $\epsilon\to0$. With respect to the
traditional
perturbative procedure where one computes first the effective action via
Feynman diagrams and then makes a variation we reversed the order: we make
first a variation of the regularized action and then compute Feynman diagrams.
One advantage of this procedure is that one never meets non-local terms which
arise typically in the traditional procedure where the anomaly stems from
diagrams with different numbers of external legs, which have to be combined
with non-local contributions, as is for example the case for non-abelian ABBJ
anomalies in any even dimension.

Let us now apply this procedure to compute the Yang-Mills anomaly coming from
the heterotic sector.
For a proper definition of the propagator for the quantum heterotic fermions
we have to augment the action
\beq
I_H={1\over2}\int d^2\sigma\gs e_p{}^i\Psib\gamma^p\gammap\Dcal_i\Psi
\hlabel{eq:hetold}\eeq
by the decoupled term
\beq
I'_H={1\over2}\int d^2\sigma
\Psib\gamma^p\gammam\partial_p\Psi
\hlabel{eq:hetlag}\eeq
which is trivially invariant under all local symmetries since we choose
$\gammam\Psi$ to be a singlet under all transformations.
The dependence on the determinant $g$ of the heterotic fermions terms
(\ref{eq:hetold}) and (\ref{eq:hetlag}) is fictitious in that $g$ can be
eliminated by rescaling the heterotic fermion fields $\Psi$. We use this
freedom to  write the heterotic fermions action as
\beq
I_H={1\over2}\int d^2\sigma\gsleft\eleft_p{}^i\Psib\gamma^p
\left(\partial_i-\gammap A_i\right)\Psi.
\hlabel{eq:hetact}\eeq
where we have introduced the left-accented zweibeins $\eleft_+{}^i=\delta_+^i$,
$\eleft_-{}^i=e_-{}^i$. Later we will use the right-accented zweibeins
$\eright_+{}^i=e_+{}^i$, $\eright_-{}^i=\delta_-{}^i$.

We have now to dimensionally extend this action; for that we shall follow the
t'Hooft-Veltman recipe as formulated by Breitenlohner and Maison
\hcite{BREMAI}.
We go to $D=2+\epsilon$
dimensions keeping consistently $\gamma^3$ strictly in two dimensions and
splitting a $D$-dimensional vector index $i$ as
$i=(\overline{\imath},\hat{\imath})$ where $\overline{\imath}$ stays strictly
in 2 and $\hat{\imath}$ denotes the extra $\epsilon$ dimensions. A similar
splitting is adopted for the flat indices $p=(\overline{p},\hat{p})$.
The Dirac algebra becomes then \hcite{BREMAI}
\beq\bgn
\{\gamma^p,\gamma^q\}&=2\eta^{pq}\nonumber\\
\{\overline{\gamma}^p,\gamma^3\}&=0\\
\left[\hat{\gamma}^p,\gamma^3\right]&=0.\nonumber
\egn\hlabel{eq:gammatau}\eeq
We compute the gauge anomaly for the classical flat metric
$\gleft^{ij}=\eta^{ij}$ restoring the metric $\gleft^{ij}$ at the end.

Performing now the transformations given in (\ref{eq:gaugevar}) we compute
the anomalous vertex associated to the dimensionally extended action gotten
from (\ref{eq:hetact}) in a flat metric to be
\beq
\Omega I_H^\epsilon={1\over2}\int d^D\sigma
\Psib\Ccal\hat{\gamma}^i\gamma^3\hat{\Dcal}_i\Psi
\eeq
where $\hat{\Dcal}_i=\hat{\partial}_i-\gammap \hat{A}_i$.
Due to the fact that the connection $A_i$ is an external field which lives
strictly in two dimensions we get for the anomalous vertex
\beq
\epsilon R_\epsilon=Q_\epsilon={1\over2}\int d^D\sigma\,
\Psib\Ccal\hat{\gamma}^i\gamma^3\hat{\partial}_i\Psi.
\hlabel{eq:anvertex}\eeq
$Q_\epsilon$ contains as external fields only the ghost field
$\Ccal=\Ccal^IT^I$ which is attached to a fermion line.

The Feynman rules are the usual ones
\beb\bno
&\mbox{{\rm $\Psi$ propagator}}\qquad{i\alpha\over\slash k}\delta^{rs}\\
&\mbox{{\rm $\Psi$-$\Psib$-$A$ gauge vertex}}
\qquad{1\over\alpha}\gamma^i{1+\gamma^3\over2}T^I.
\eno
The Feynman rule associated to the anomalous vertex (\ref{eq:anvertex}) is
given by
\beq
\mbox{{\rm $\Psi$-$\Psi$-$\Ccal$ anomalous vertex}}
\qquad{i\over\alpha}{\hat{\slash k}-\hat{\slash k}'\over2}\gamma^3T^J
\hlabel{eq:franvertex}\eeq\eeb
where $J$ is the gauge index carried by the external ghost field and $k$ and
$k'$ are the incoming and outgoing momenta of the fermions. The Feynman
graphs at one loop with the insertion of {\it one\/} anomalous vertex of the
type (\ref{eq:franvertex}) are indicated in Fig.~\ref{fig:anom}.

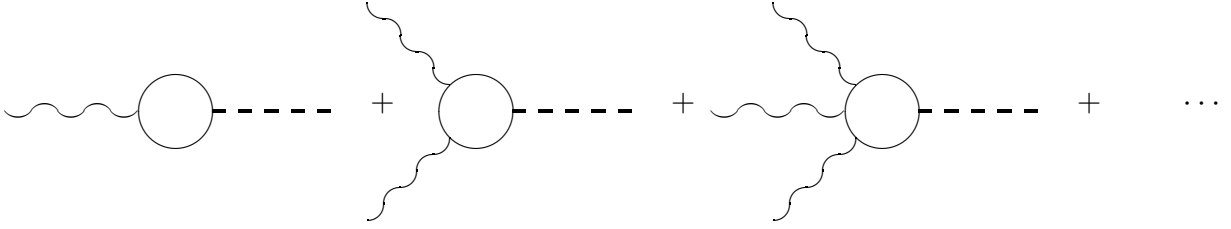
\begin{figure}[t] 
\begin{picture}(28000,8000)(-5000,0)
\pfrontx=1000 \pfronty=4000
\put(\pfrontx,\pfronty){\circle{2800}}
\global\advance\pfrontx by -1500
\drawline\photon[\W\REG](\pfrontx,\pfronty)[5]
\global\advance\pfrontx by 2870
\thicklines
\multiput(\pfrontx,\pfronty)(1000,0){5}{\line(1,0){500}}
\thinlines
\global\advance\pfrontx by 6000
\put(\pfrontx,\pfronty){$+$}
\global\advance\pfrontx by 4000
\put(\pfrontx,\pfronty){\circle{2800}}
\global\advance\pfrontx by -1000
\global\advance\pfronty by 1000
\drawline\photon[\NW\REG](\pfrontx,\pfronty)[5]
\advance\pfronty by -2000
\drawline\photon[\SW\REG](\pfrontx,\pfronty)[5]
\global\advance\pfrontx by 2370
\global\advance\pfronty by 1000
\thicklines
\multiput(\pfrontx,\pfronty)(1000,0){5}{\line(1,0){500}}
\thinlines
\global\advance\pfrontx by 6000
\put(\pfrontx,\pfronty){$+$}
\global\advance\pfrontx by 8000
\put(\pfrontx,\pfronty){\circle{2800}}
\global\advance\pfrontx by -1000
\global\advance\pfronty by 1000
\drawline\photon[\NW\REG](\pfrontx,\pfronty)[5]
\global\advance\pfronty by -2000
\drawline\photon[\SW\REG](\pfrontx,\pfronty)[5]
\global\advance\pfrontx by 2000

\global\advance\pfrontx by -2500
\global\advance\pfronty by 1000
\drawline\photon[\W\REG](\pfrontx,\pfronty)[5]
\global\advance\pfrontx by 2870

\thicklines
\multiput(\pfrontx,\pfronty)(1000,0){5}{\line(1,0){500}}
\thinlines
\global\advance\pfrontx by 6000
\put(\pfrontx,\pfronty){$+$}
\global\advance\pfrontx by 4000
% ...
\put(\pfrontx,\pfronty){$\cdots$}
\end{picture}
\caption{Fermionic graphs contributing to the one-loop anomaly.}
\tlabel{fig:anom}
\end{figure}

Let us compute the anomaly arising from the first diagram in that figure; it
contains one external gauge field $A^I$ and an external ghost $C^J$ associated
to (\ref{eq:franvertex}) while in the loop are circulating fermions. Keeping
the
external momenta strictly in two dimensions it is given by
\beq
{i\over\alpha}{\cal A}_j^{IJ}(p)
={i\over2}\trace(T^IT^J)\int{d^Dk\over(2\pi)^D}\trace
\left(\hat{\slash k}\gamma_3{1\over\slash k}
\gamma_j{1+\gamma_3\over2}{1\over\slash k-\slash p}\right).
\hlabel{eq:69tex}\eeq
The integral over $k$ is ultraviolet (logarithmically) divergent, on the other
hand $\hat{k}$ is of order $\epsilon$ so that the result is expected to be
finite. A careful calculation gives in fact, in the limit $\epsilon\to0$:
\beq
{\cal A}_j^{IJ}={\alpha\over8\pi}\trace(T^IT^J)
(\eta_{mj}-\varepsilon_{mj})(ip^m).
\eeq
Upon adding the external legs, $A^I$ and ${\cal C}^J$, and
transforming back to configuration space one gets for the gauge anomaly
\beq
{\cal A}'_G=-{\alpha\over8\pi}\int d^2\sigma\trace({\cal C}\partial_+A_-).
\hlabel{eq:1lanom}\eeq
If we go on to consider the diagrams with $n$ external legs in
Fig.~\ref{fig:anom}, we may notice that the integration over the loop-momentum
behaves for large $k$ as
\beq
\sim\int^\Lambda {d^Dk\over k^{n+1}}\hat{k}\sim\Lambda^{2-n}.
\hlabel{eq:sim}\eeq
Now for $n\geq3$ the integral over $k$ in (\ref{eq:sim}) is surely convergent
and due to the presence of $\hat{k}$ in the numerator, as
$\epsilon\to0$ the amplitude vanishes. So there is no contribution to the
anomaly coming from all the diagrams  in Fig.~\ref{fig:anom} with three or
more external gauge fields. The unique case to be considered remains the
diagram with the insertion of {\it two\/} gauge fields. In this case one gets
a logarithmically divergent integral in (\ref{eq:sim}) and for $\epsilon\to0$
one can set the external momenta to zero. For the second diagram in
Fig.~\ref{fig:anom} one gets
\beq
\Acal_2=\int{d^Dk\over(k^2)^3}\Jcal(k^3,\hat{k}).
\eeq
The function $\Jcal(k^3,\hat{k})$ is written explicitly in the appendix. It is
constituted by a trace over $\gamma$-matrices containing
three powers of momenta in $D$ dimensions and one $\hat{k}$ which lives in
$\epsilon$ dimensions. A careful analysis of this trace of $\gamma$-matrices
reveals, however, that actually $\Acal_2=0$ identically (see appendix A).
As a result also this diagram vanishes and we are left with
the anomaly computed in (\ref{eq:1lanom}) .

Restoring the left-accented metric we have
\beq
\Acal'_G=-{\alpha\over8\pi}\int d^2\sigma\gsleft
\trace(\Ccal\Dleft_+A_-),
\hlabel{eq:71tex}\eeq
which is not invariant under diffeomorphisms.
Here we defined the Weyl and $d=2$ local Lorentz covariant derivatives for a
generic zweibein $e_\pm{}^i$
\beq
D_\pm=\partial_\pm+{1\over\gs}\partial_j(\gs e_\pm{}^j).
\hlabel{eq:99}\eeq
We can get a diff-invariant form
of the anomaly by adding a local term (in dimensional regularization the
effective action is always defined modulo local terms); we redefine the
effective action according to
\beq
\Gamma_H=\Gamma'_H-{\alpha\over16\pi}\int d^2\sigma\gsleft\trace(\Aleft_+A_-)
\hlabel{eq:cocycle}\eeq
to get the metric-independent gauge anomaly
\beq
\Acal_G={\alpha\over8\pi}\int d^2\sigma\varepsilon^{ij}
\trace(\Ccal\partial_iA_j).
\hlabel{eq:gauanom}\eeq
Clearly this anomaly can also be deduced directly by integrating
(\ref{eq:hetact}) over the fermions and computing the $A$-$A$ contribution to
the effective action (Fig.~\ref{fig:eafermions}). For a flat metric, with our
(dimensional) regularization, one gets
\bno
\Gamma'_H&=-{\alpha\over16\pi}\int d^2\sigma
(\eta^{ij}-\varepsilon^{ij})(\eta^{mn}-\varepsilon^{mn})
\trace\left(\partial_iA_j{1\over\Box}\partial_mA_n\right)\\
&=-{\alpha\over16\pi}\int d^2\sigma
\trace\left(\partial_+A_-{1\over\Box}\partial_+A_-\right)\nonumber.
\eno
We can restore the left-accented metric to obtain
\beq
\Gamma'_H=-{\alpha\over16\pi}\int d^2\sigma\gsleft\trace
\left(\Dleft_+A_-{1\over\Box_\gleft}\Dleft_+A_-\right),
\hlabel{eq:eametr}\eeq
where, for a generic metric $g_{ij}$,
\beq
\Box_g\equiv{1\over\gs}\partial_i(\gs g^{ij}\partial_j)
=D_+\partial_-=D_-\partial_+.
\eeq
As it stands, (\ref{eq:eametr}) suffers a diffeomorphisms anomaly which is
however trivial and can be eliminated by redefining $\Gamma'_H$ as in
(\ref{eq:cocycle}) to get finally:
\bns
\Gamma_H&=-{\alpha\over16\pi}\int d^2\sigma\gsleft\trace
\left(\Dleft_+A_-{1\over\Box_\gleft}(\Dleft_+A_--D_-\Aleft_+)\right)\\
&=-{\alpha\over8\pi}\int d^2\sigma\gs\trace
\left(A_-{1\over D_-}{\varepsilon^{ij}\partial_iA_j\over\gs}\right).\yesnumber
\hlabel{eq:ea100}\ens
It is not difficult to convince ourselves that actually the determinant $\gs$
scales away in (\ref{eq:ea100}) and therefore we were allowed to replace
$\gsleft$ with $\gs$.
Varying this action according to (\ref{eq:gaugevar}) we get
\bno
\delta_G\Gamma_H&={\alpha\over8\pi}\int d^2\sigma\varepsilon^{ij}
\trace({\cal C}\partial_iA_j)-\nonumber\\
&\qquad-{\alpha\over8\pi}\int d^2\sigma\gs
\trace\left(D_+[\Ccal,A_-]{1\over \Box_g}D_+A_-\right).
\hlabel{eq:eavar}\eno
The first term in (\ref{eq:eavar}) is local and corresponds to the anomaly
(\ref{eq:gauanom}) while
the second term is non-local and is clearly spurious in the sense that it gets
cancelled by a corresponding term in the variation of $\Gamma_3$, see the
second diagram in Fig.~\ref{fig:eafermions} with three external gauge fields
$A_i$. Now, also $\delta(\Gamma_H+\Gamma_3)$
contains, apart from ${\cal A}_G$, non-local terms which are cancelled by
$\delta\Gamma_4$ and so on. These cumbersome linked cancellations which are
due to the non abelian nature of the Yang-Mills gauge fields are elegantly
avoided by the non-standard method we employed above, because in that case
the diagrams with two or more external gauge fields do simply not contribute.
Actually, our non-standard derivation of the gauge anomaly constitutes a proof
of these linked cancellations.

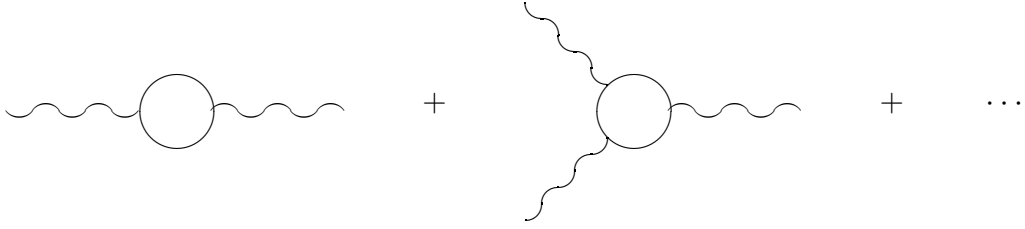
\begin{figure}[t] 
\begin{picture}(28000,8000)(-8000,0)
\pfrontx=1000 \pfronty=4000
\put(\pfrontx,\pfronty){\circle{2800}}
\global\advance\pfrontx by -1500
\drawline\photon[\W\REG](\pfrontx,\pfronty)[5]
\global\advance\pfrontx by 2800
\drawline\photon[\E\REG](\pfrontx,\pfronty)[5]
\global\advance\pfrontx by 8000
\put(\pfrontx,\pfronty){$+$}
\global\advance\pfrontx by 8000
\put(\pfrontx,\pfronty){\circle{2800}}
\global\advance\pfrontx by -1000
\global\advance\pfronty by 1000
\drawline\photon[\NW\REG](\pfrontx,\pfronty)[5]
\advance\pfronty by -2000
\drawline\photon[\SW\REG](\pfrontx,\pfronty)[5]
\global\advance\pfrontx by 1000
\global\advance\pfrontx by 1300
\global\advance\pfronty by 1000
\drawline\photon[\E\REG](\pfrontx,\pfronty)[5]
\global\advance\pfrontx by 8000
\put(\pfrontx,\pfronty){$+$}
\global\advance\pfrontx by 4000
% ...
\put(\pfrontx,\pfronty){$\cdots$}
\end{picture}
\caption{Fermionic graphs contributing to the one-loop effective action.}
\tlabel{fig:eafermions}
\end{figure}

We turn now to the derivation of the $\kappa$-anomaly in the Yang-Mills
sector coming from the functional integration over the heterotic fermions of
(\ref{eq:hetact}). The $\kappa$-transformations of the fields comparing in
(\ref{eq:hetact}) are given by (here we use again the two-component notation)
\beb\bns
\delta_\kappa A_i&=\Dcal_iC+F_i\yesnumber\\
\delta_\kappa\Psi&=\gammap\Big[C\Psi+\big(\Delta^\alpha y^bF_{b\alpha}
+{1\over2}y^CD_C\Delta^\alpha y^bF_{b\alpha}\\
&+{1\over2}\Delta^\alpha y^by^C\Dcal_CF_{b\alpha}+o(y^3)\big)
(\psi_0+\Psi)\Big].\yesnumber\\
\delta_\kappa\eleft_p{}^i&=0
\ens\eeb
where we recall that $C=\Delta^\alpha A_\alpha$. As we observed already,
the $\kappa$-transformations act like a field-dependent gauge
transformation with parameter $C$ plus an intrinsic $\kappa$-transformation.
Notice that $F_-=0$, see (\ref{eq:FRdual}), and that only $A_-$ is coupled to
the heterotic fermions in (\ref{eq:hetact}).

The field-dependent gauge transformation gives therefore rise to an anomalous
$\kappa$-vertex which is given by (\ref{eq:anvertex}) where $\Ccal$ has to be
substituted by $C$. The related $\kappa$-anomaly can then be computed in
complete analogy to the gauge anomaly (\ref{eq:71tex}) and one gets
\beq
\Acal_G^\kappa=-{\alpha\over8\pi}\int d^2\sigma\gsleft\trace
\left(C\Dleft_+A_-\right).
\hlabel{eq:gaukanom}\eeq
Again this anomaly is not diff-invariant and we add to the effective
action the same cocycle as in (\ref{eq:cocycle}) to obtain the diff-invariant
$\kappa$-anomaly,
\bno
\Acal_G^\kappa&={\alpha\over8\pi}\int d^2\sigma\varepsilon^{ij}
\trace(C\partial_iA_j+F_iA_j)\nonumber\\
&=-{\alpha\over16\pi}\int d^2\sigma\varepsilon^{ij}
V_i{}^AV_j{}^B\Delta^\gamma(\omega_{3YM})_{\gamma BA}.
\hlabel{eq:55}\eno
The first line in (\ref{eq:55}) stems from (\ref{eq:FRdual}) while the
second line involves the definition of the Yang-Mills Chern-Simons form
$\omega_{3YM}$ given in (\ref{eq:csdef}). The intrinsic
$\kappa$-transformations are expected not to contribute to the
$\kappa$-anomaly at one loop since, as we will see in section VIII,
(\ref{eq:55}) satisfies already the Wess-Zumino consistency condition.

Taking a look at (\ref{eq:actkvar}) one realizes that ${\cal A}_G^\kappa$ can
be eliminated \hcite{ATICK} by imposing the constraints, which are imposed on
$W=dB$ in (\ref{eq:grconst}) at the classical level, on the three-superform
$H$ defined as
\beq
H=dB+{\alpha\over8\pi}\omega_{3YM}.
\hlabel{eq:Hdef}\eeq
This relation then requires that $B$ has to transform anomalously under
gauge-transformations according to
\beq
\delta_GB=-{\alpha\over8\pi}\trace({\cal C}dA)
\hlabel{eq:Bvar}\eeq
because $\delta_G\omega_{3YM}=d({\cal C}dA)$. Then, taking (\ref{eq:Bvar})
into account, the gauge transformation of the action (\ref{eq:action}) cancels
the gauge anomaly (\ref{eq:gauanom}), as is well known.

The Bianchi identity associated to (\ref{eq:Hdef}) is
\beq
dH={\alpha\over8\pi}\trace(FF),
\hlabel{eq:84tex}\eeq
it can be consistently solved in superspace \hcite{ATIRAT}, and it gives rise
to the Chapline-Manton theory \hcite{CHAMAN}, i.e. constitutes the minimally
coupled SUGRA-SYM theory in ten dimensions. Eq.~(\ref{eq:84tex}) coincides
with the result of Ref.~\hcite{GRIZAN} by taking into account that our $H$
differs from the one used in that reference by a factor of two.
\vspace{1truecm}
\section{The Lorentz anomaly}
In this section we want to derive the Lorentz anomaly of the sigma model with
the same technique we used in the previous section to derive the gauge
anomaly. A Lorentz anomaly is expected to appear due to the chiral coupling of
the anticommuting $y^\alpha$ to the induced Lorentz connection
$\Omega_{i\alpha}{}^\beta\equiv{1\over4}\Omega_{iab}
(\Gamma^{ab})_\alpha{}^\beta$, $\Omega_{ia}{}^b\equiv V_i{}^C\Omega_{Ca}{}^b$,
in the first term in (\ref{eq:fullexp}), through the covariant derivative
$D_jy^\beta\equiv\partial_jy^\beta-\Omega_j{}^\beta{}_\gamma y^\gamma$. This
term is invariant under the Lorentz transformations
\beb\bno
\delta_Ly^\alpha&=\Lcal^\alpha{}_\beta y^\beta\\
\delta_L\Omega_i{}^\beta{}_\alpha&=\partial_i\Lcal^\beta{}_\alpha
+\Lcal^\beta{}_\gamma\Omega_i{}^\gamma{}_\alpha
-\Omega_i{}^\beta{}_\gamma\Lcal^\gamma{}_\alpha
\hlabel{eq:104}\\[5pt]
\Lcal^\beta{}_\alpha&={1\over4}(\Gamma_{ab})^\beta{}_\alpha\Lcal^{ab}.\nonumber
\eno\eeb
Since there exists up to now no $SO(10)$ Lorentz-covariant quantization of the
theory we
limit ourselves to derive the Lorentz anomaly under $SO(8)$ transformations,
i.e. such that
\beq
m_a\Lcal^{ab}=n_a\Lcal^{ab}=0.
\hlabel{eq:105}
\eeq
To get a canonical kinetic term for the worldsheet scalars $y^\alpha$ they
have to be transformed to worldsheet Majorana-Weyl fermions \hcite{GSBOOK}.
This can be achieved by rescaling the $y^\alpha$ by an $SO(8)$ invariant
quantity
\beq
y^\alpha={1\over\sqrt{4n_-}}y_u^\alpha,
\eeq
where $n_-=V_-{}^an_a$, and by introducing a worldsheet Majorana spinor as
\beq
Y^\alpha=\left(\begin{array}{c}y_u^\alpha\\y_d^\alpha\end{array}\right)
\eeq
whose bottom component $1/2(1+\gamma^3)Y^\alpha=\left(\begin{array}{c}0\\
y_d^\alpha\end{array}\right)$ is decoupled from the theory.

Then the first term in (\ref{eq:fullexp}) can be rewritten as follows:
\beq
\gs y^\alpha \Vsla_{-\alpha\beta}D_+y^\beta
={1\over4n_-}\gs
\Ybar^\alpha e_p{}^i\gamma^p\gammam\Vsla_{-\alpha\beta}D_iY^\beta.
\eeq
To complete the action of the Majorana-Weyl fermions $Y^\alpha$ we have to add
the decoupled kinetic term of the $y_d^\alpha$; in analogy with the
discussion on the heterotic fermions kinetic term we get
\beq
I_F={1\over2}\int d^2\sigma\gsright\left({1\over2n_-}
\Ybar^\alpha \eright_p{}^i\gamma^p\gammam\Vsla_{-\alpha\beta}D_iY^\beta
+\eright_p{}^iY^\alpha\gamma^p\gammap\msla_{\alpha\beta}
\partial_iY^\beta\right).
\hlabel{eq:107}\eeq
where now we use the right-accented zweibeins $\eright_+{}^i=e_+{}^i$,
$\eright_-{}^i=\delta_-^i$.
$I_F$ is invariant under $SO(8)$ local transformations, i.e. (\ref{eq:104}) and
\beq
\delta_LY^\alpha=\gammam\Lcal^\alpha{}_\beta Y^\beta.
\eeq
Now we can proceed along the lines of the preceding section to compute the
Lorentz anomaly. We use dimensional regularization to extend $I_F$ to
$I_F^\epsilon$ in precisely the same manner as we did in the preceding section
for the heterotic fermions and compute the anomalous vertex associated to
(\ref{eq:107}) for a flat metric $\gright^{ij}=\eta^{ij}$. We get
\beq
\Omega_LI_F^\epsilon=-{1\over2}\int d^D\sigma
\Ybar^\alpha\Lcal_\alpha{}^\beta\gammah^i\gamma^3
\left({V_-{}^a\over2n_-}(\Gamma_a)_{\beta\gamma}\gammam D_i
+\msla_{\beta\gamma}\gammap\partial_i\right)Y^\gamma.
\hlabel{eq:110}\eeq
To compute the anomaly we enforce now the $\kappa$ gauge-fixing (\ref{eq:gfix})
which becomes
\beq
\gammam (\nsla Y)_\alpha=0.
\eeq
To do this we insert the identity $\nsla\msla+\msla\nsla=1$ in (\ref{eq:107})
and in (\ref{eq:110}), to get
\bno
\Itil_F^\epsilon&={1\over2}\int d^D\sigma\Ybar^\alpha\gamma^i
\left(\partial_i\msla_{\alpha\beta}-\gammam\Omega_{i\alpha}{}^\gamma
\msla_{\gamma\beta}\right)Y^\beta\hlabel{eq:112}\\
\epsilon R_\epsilon&=Q_\epsilon=-{1\over2}\int d^D\sigma
Y^\alpha\Lcal_\alpha{}^\beta\gammah^i\gamma^3
\msla_{\beta\gamma}\partialh_iY^\gamma.\hlabel{eq:113}
\eno
We used the fact that $[\Lcal,\nsla]=0=[\Omega_i,\nsla]$ and that
$\Omega_{i\alpha}{}^\beta$ lives strictly in two dimensions.

Now we can use the formal analogy between (\ref{eq:112}), (\ref{eq:113}) and
(\ref{eq:hetact}), (\ref{eq:anvertex})  to compute the anomaly. From
(\ref{eq:112}) we deduce the Feynman rules
\beb\bno
&\mbox{$Y^\alpha$ propagator}\qquad{i\alpha\over\ksla}\nsla\\
&\mbox{$Y$-$Y$-$\Omega$ vertex}\qquad{1\over\alpha}\gamma^i\gammam{1\over4}
(\Gamma_{ab}\msla)_{\alpha\beta}
\eno
while for the anomalous vertex we get from (\ref{eq:113}) the Feynman rule
\beq
\mbox{$Y$-$Y$-$\Lcal$ anomalous vertex}
\qquad-{i\over\alpha}{\hat{\ksla}-\hat{\ksla}'\over2}\gamma^3
{1\over4}(\Gamma_{cd}\msla)_{\alpha\beta}.
\hlabel{eq:114}\eeq\eeb
The anomaly can now be computed in the same way as in the preceding section,
one only has to flip the chiralities. The first diagram in Fig.~\ref{fig:anom}
with
the insertion of the anomalous vertex (\ref{eq:114}) and one external
$\Omega_{jab}$ gives
\beq
{i\over\alpha}\Acal^{cd}_{jab}(p)={i\over32}\trace
\left(\Gamma^{cd}\msla\nsla\Gamma_{ab}\msla\nsla\right)
\int{d^Dk\over(2\pi)^D}\trace
\left(\hat{\ksla}\gamma^3{1\over\ksla}\gamma_j
\gammam{1\over\ksla-\psla}\right).
\hlabel{eq:115}\eeq
The integral in (\ref{eq:115}) has already been calculated in the previous
section (see (\ref{eq:69tex}))  while the trace of $\Gamma$-matrices, apart
from terms which go to
zero due to (\ref{eq:ofix}), can be calculated to give
\beq
\trace\left(\Gamma^{cd}\msla\nsla\Gamma_{ab}\msla\nsla\right)
=-16\delta_{[a}^c\delta_{b]}^d.
\hlabel{eq:115a}\eeq
The result for $\epsilon\to0$ is
\beq
\Acal_{jab}^{cd}=-{\alpha\over8\pi}\delta_{[a}^c\delta_{b]}^d
(\eta_{mj}+\varepsilon_{mj})(ip^m).
\eeq
Adding the external legs $\Lcal_{cd}$ and $\Omega_j{}^{ab}$ and restoring the
right-accented zweibeins we get for the $SO(8)$ Lorentz anomaly
\beq
\Acal'_L=-{\alpha\over8\pi}\int d^2\sigma\gsright\trace
\left(\Lcal \Dright_-\Omega_+\right).
\hlabel{eq:116}\eeq
Here the traces are in the fundamental representation of the Lorentz group,
$\trace(\Lcal\Omega_j)\equiv\Lcal_{ab}\Omega_j{}^{ba}$. Eq.~(\ref{eq:116})
gives the anomaly under $SO(8)$ transformations. We postulate that the anomaly
under $SO(10)$ transformations, in an eventual covariant
quantization scheme, is still given by (\ref{eq:116}) where the constraints
(\ref{eq:ofix}) and (\ref{eq:105}) are released. Also in this case the
diagrams with two or more external $\Omega_i$ fields and the insertion of an
anomalous vertex are zero for $\epsilon\to0$.

Again, to render the anomaly diff-invariant we add a trivial cocycle to the
effective action as in (\ref{eq:cocycle})
\beq
\Gamma_F=\Gamma'_F-{\alpha\over16\pi}\int d^2\sigma\gsright
\trace\left(\Omegar_-\Omega_+\right),
\hlabel{eq:cocycle2}\eeq
so that
\beq
\Acal_L=-{\alpha\over8\pi}\int d^2\sigma\varepsilon^{ij}
\trace\left(\Lcal\partial_i\Omega_j\right).
\hlabel{eq:120}\eeq
The direct computation of the Lorentz anomaly in this case requires to compute
the $\Omega$-$\Omega$ contribution to the effective action coming from the
integration over the fermions $Y^\alpha$ in (\ref{eq:112}). The computation is
standard, all one has to use is again (\ref{eq:115a}) and the result is
completely analogous to (\ref{eq:eametr}):
\beq
\Gamma'_F=-{\alpha\over16\pi}\int d^2\sigma\gsright\trace
\left(\Dright_-\Omega_+{1\over\Box_\gright}\Dright_-\Omega_+\right).
\hlabel{eq:117}\eeq
By adding the trivial cocycle as in (\ref{eq:cocycle2}) we get an expression
analogous to (\ref{eq:ea100}),
\beq
\Gamma_F={\alpha\over8\pi}\int d^2\sigma\gs\trace
\left(\Omega_+{1\over D_+}{\varepsilon^{ij}\partial_i\Omega_j\over\gs}\right)
\hlabel{eq:104tex}\eeq
which is now diff-invariant. Its Lorentz variation is
\bno
\delta_L\Gamma_F&=-{\alpha\over8\pi}\int d^2\sigma\varepsilon^{ij}\trace
\left(\Lcal \partial_i\Omega_j\right)-\nonumber\\
&\qquad-{\alpha\over8\pi}\int d^2\sigma\gs\trace
\left(D_-[\Lcal,\Omega_+]{1\over\Box_g}D_-\Omega_+\right).
\hlabel{eq:118}\eno
Again, the first line is the anomaly (\ref{eq:120}) while the second line is
non-local and gets cancelled by the diagram with three external $\Omega$'s,
see Fig.~\ref{fig:anom}.

For a first attempt on the derivation of Eq.~(\ref{eq:118}) see \hcite{HAAG}.
Let us briefly discuss the appearance of additional Lorentz anomalies.
Generally speaking they can arise from the terms in (\ref{eq:fullexp}) where
the connection $\Omega_i$ appears explicitly. In the term
$-{1\over2}\gs g^{ij}D_iy^aD_jy_a$ the connection is non-chirally coupled, so
no Lorentz anomaly can arise. For what concerns the mixed term
$-2\gs D_-y^aV_+{}^\alpha(\Gamma_a)_{\alpha\beta}y^\beta$, to preserve
manifest $SO(8)$ invariance we have to impose the physical condition on the
external field $V_j{}^\beta$, $\nsla_{\alpha\beta}V_j{}^\beta=0$. Then upon
inserting the identity $\nsla\msla+\msla\nsla=1$, this term becomes
$-4\gs\partial_-y^an_aV_+{}^\beta\msla_{\beta\alpha}y^\alpha$ such
that the connection drops due to (\ref{eq:ofix}). The seventh term in
(\ref{eq:fullexp}) contains $\Omega_i$ explicitly but does not contribute to
the Lorentz anomaly as we will see in the next section.

The terms which are quadratic in the $y^\alpha$ in (\ref{eq:fullexp}) give
rise to ``trivial'' anomalies and do therefore not constitute ``anomalies''.
We evidenciate this fact for the nineth term. To preserve $SO(8)$ invariance
we have to impose on $T_{abc}$ the condition
\beq
n_aT^{abc}=0=m_aT^{abc}.
\eeq
Then this term can be taken into account simply by defining
\beq
\Omegat_{ia}{}^b\equiv\Omega_{ia}{}^b-{1\over2}e_{-i}V_+{}^gT_{ga}{}^b,
\eeq
that is:
\beq\bgn
\Omegat_{+a}{}^b&=\Omega_{+a}{}^b-V_+{}^gT_{ga}{}^b\\
\Omegat_{-a}{}^b&=\Omega_{-a}{}^b.
\egn\eeq
This would produce instead of (\ref{eq:116}) the anomaly
\bns
\Acalt_L&=-{\alpha\over8\pi}\int d^2\sigma\gsright\trace
\left(\Lcal \Dright_-\Omegat_+\right)\\
&=\Acal'_L-{\alpha\over8\pi}\int d^2\sigma\gsright\,
\partialr_-\Lcal_{ab}V_+{}^cT_c{}^{ba}\\
&=\Acal'_L+\delta_L\left(-{\alpha\over8\pi}\int d^2\sigma\gsright\,
\Omegar_{-ab}V_+{}^cT_c{}^{ba}\right)\yesnumber\hlabel{eq:108tex}
\ens
and therefore $\Acalt_L$ and $\Acal_L$ represent the same cohomology
class.

The Lorentz anomaly can be cancelled if we subject the two-superform $B$ to
the anomalous Lorentz transformation
\beq
\delta_LB={\alpha\over8\pi}\trace(\Lcal d\Omega)
\eeq
which, together with (\ref{eq:Bvar}), defines the gauge and Lorentz invariant
curvature
\beq
H=dB+{\alpha\over8\pi}(\omega_{3YM}-\omega_{3L})
\hlabel{eq:122}\eeq
with the associated Bianchi identity in superspace
\beq
dH={\alpha\over8\pi}\left(\trace F^2-\trace R^2\right).
\hlabel{eq:123}\eeq
Notice that both traces in (\ref{eq:123}) are in the fundamental
representations of $SO(32)$ and $SO(10)$ respectively and, according to the
Green-Schwarz anomaly cancellation mechanism, this is then also precisely the
relation which assures the absence of gauge and Lorentz anomalies in $N=1$,
$D=10$ Supergravity-Super-Yang-Mills theory.

In the next section we will show that (\ref{eq:122}), (\ref{eq:123}) are
actually sufficient and necessary to cancel also the Lorentz $\kappa$-anomaly
in our sigma model.
\vspace{1truecm}
\section{The Lorentz-type $\kappa$-anomaly}
At this point an important difference between the gauge sector and the
gravitational sector shows up. The gauge-type $\kappa$-anomaly could be
calculated by simply varying (\ref{eq:eametr}) while the Lorentz-type
$\kappa$-anomaly can not be computed by varying simply (\ref{eq:117}). This
can easily be seen by observing that in (\ref{eq:117}) with respect to
(\ref{eq:eametr}) the chiralities are flipped. For the
$\kappa$-transformations of the induced connections we have
\bns
\delta_\kappa A_i&=\Dcal_iC+F_i\\
\delta_\kappa\Omega_{ia}{}^b&=D_iL_a{}^b+R_{ia}{}^b
\ens
where in both cases, see (\ref{eq:FRdual}), $F_-=0$ and $R_{-a}{}^b=0$,
while $F_+$ and $R_{+a}{}^b$ are different from
zero. Therefore the variation of $\Gamma_F$ gives, unlike as in the Yang-Mills
case, apart from a local contribution, non-local contributions proportional to
$R_+$; moreover $\Gamma_F$ depends non-locally on $e_+{}^j$ and the
$\kappa$-variation of $e_+{}^j$ induces additional non-local terms. It can
also be seen that the local terms in $\delta_\kappa\Gamma_F$ do not satisfy
the Wess-Zumino consistency condition, see the next section.

The key observation for the resolution of this puzzle is that, as can be seen
from (\ref{eq:kyvar}), $\kappa$-transformations mix the fermions $y^\alpha$
with the bosons $y^a$. The Lorentz-type $\kappa$-anomaly stems from the
explicit coupling of the induced Lorentz-connection $\Omega_i$ to the quantum
fields $(y^a,y^\alpha)$. While the $y^a$ do not contribute to the
Lorentz-anomaly, as we mentioned already, they are expected to contribute to
the Lorentz-type $\kappa$-anomaly because of their explicit coupling to the
$\Omega_i$ in the term $-{1\over2}\gs g^{ij}D_iy^aD_jy_a$. Their contribution
is actually essential to saturate the coupled cohomology problem
(\ref{eq:coupled2}). The analogy with the supersymmetric partner of an ABBJ
anomaly in a $d=2$ Super-Yang-Mills theory has already been discussed in the
introduction.

Since massless scalars in two dimensions, as are the $y^a$, are always plagued
by infrared divergences we introduce an infrared mass regulator $m$ and take
the relevant boson action to be
\beq
I_B=-{1\over2}\int d^2\sigma\gs
\left(g^{ij}D_iy^aD_jy_a-m^2y^ay_a\right).
\hlabel{eq:124}\eeq
Remember that $D_iy^a=\partial_iy^a+y^b\Omega_{ib}{}^a$. The Feynman rules for
$g^{ij}=\eta^{ij}$ are
\beb\bno
&\mbox{{\rm $y^a$ propagator}}\qquad-{i\alpha\over k^2-m^2}\eta_{ab}\\
&\mbox{{\rm$\Omega$-$y$-$y$ vertex}}\qquad{1\over\alpha}(k+k')i
\delta_{[a}^c\delta_{b]}^d\\
&\mbox{{\rm$\Omega$-$\Omega$-$y$-$y$ vertex}}\qquad-{2i\over\alpha}
\eta^{ij}\eta_{bd}\eta_{fa}\eta_{gc}.
\eno\eeb
The last vertex has to be saturated with the external legs
$\Omega_i{}^{ab}\Omega_j{}^{cd}$ while $f$ and $g$ indicate the internal boson
lines.

\begin{figure}[t] 
\begin{picture}(28000,8000)(-5000,0) 
\pfrontx=1000 \pfronty=4000
\put(\pfrontx,\pfronty){\circle{2800}}
\put(\pfrontx,\pfronty){\circle{2600}}
\global\advance\pfrontx by -1500
\drawline\photon[\W\REG](\pfrontx,\pfronty)[5]
\global\advance\pfrontx by 2800
\drawline\photon[\E\REG](\pfrontx,\pfronty)[5]
\global\advance\pfrontx by 8000
\put(\pfrontx,\pfronty){$+$}
\global\advance\pfrontx by 8000
\put(\pfrontx,\pfronty){\circle{2800}}
\put(\pfrontx,\pfronty){\circle{2600}}
\global\advance\pfrontx by -1500
\drawline\photon[\NW\REG](\pfrontx,\pfronty)[5]
\global\advance\pfronty by 300
\drawline\photon[\SW\REG](\pfrontx,\pfronty)[5]
\global\advance\pfronty by -300
\global\advance\pfrontx by 2800
\global\advance\pfrontx by 8000
\put(\pfrontx,\pfronty){$+$}
\global\advance\pfrontx by 4000
% ...
\put(\pfrontx,\pfronty){$\cdots$}
\end{picture}

\begin{picture}(28000,8000)(-5000,0) 
\pfrontx=1000 \pfronty=4000
\put(\pfrontx,\pfronty){\circle{2800}}
\put(\pfrontx,\pfronty){\circle{2600}}
\global\advance\pfrontx by -1000
\global\advance\pfronty by 1000
\drawline\photon[\NW\REG](\pfrontx,\pfronty)[5]
\advance\pfronty by -2000
\drawline\photon[\SW\REG](\pfrontx,\pfronty)[5]
\global\advance\pfrontx by 2300
\global\advance\pfronty by 1000
\drawline\photon[\E\REG](\pfrontx,\pfronty)[5]
\global\advance\pfrontx by 8000
\put(\pfrontx,\pfronty){$+$}
\global\advance\pfrontx by 8000
\put(\pfrontx,\pfronty){\circle{2800}}
\put(\pfrontx,\pfronty){\circle{2600}}
\global\advance\pfrontx by -1500
\drawline\photon[\NW\REG](\pfrontx,\pfronty)[5]
\global\advance\pfronty by 300
\drawline\photon[\SW\REG](\pfrontx,\pfronty)[5]
\global\advance\pfronty by -300
\global\advance\pfrontx by 2800
\drawline\photon[\E\REG](\pfrontx,\pfronty)[5]
\global\advance\pfrontx by 8000
\put(\pfrontx,\pfronty){$+$}
\global\advance\pfrontx by 4000
% ...
\put(\pfrontx,\pfronty){$\cdots$}
\end{picture}
\caption{Bosonic graphs contributing to the one-loop effective action.}
\tlabel{fig:eabosons}
\end{figure}
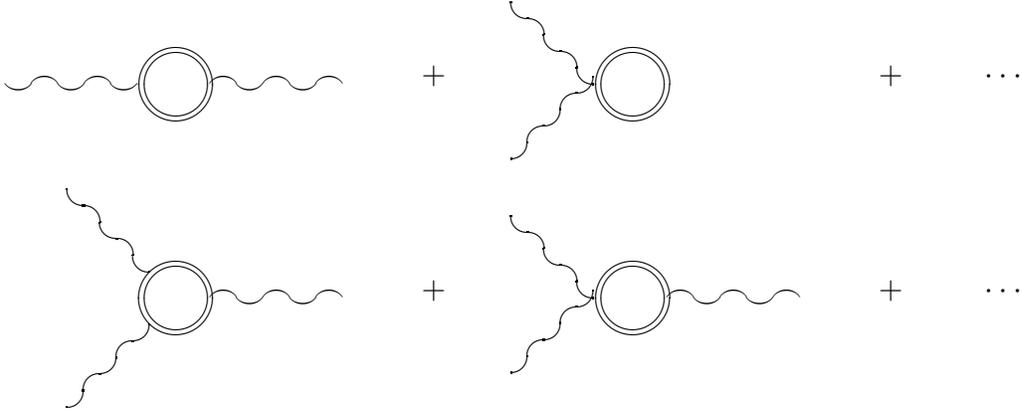

We compute the contribution of (\ref{eq:124}) to the effective action which
is quadratic in the $\Omega_i$. We have a self-energy type diagram and a
tadpole diagram (the first two pictures in Fig.~\ref{fig:eabosons}). Since
each of the two diagrams is individually ultraviolet divergent we introduce
also here a dimensional regularization with $D=2+\epsilon$ and a scale $\mu$ to
compute them. Adding up the two diagrams we get in momentum space for generic
$m$ and $\epsilon$
\beq
{1\over\alpha}\Gamma_{iab;j}{}^{cd}(p)=\delta_{[a}^d\delta_{b]}^c
\left(\eta_{ij}-{p_ip_j\over p^2}\right)B(p^2)
\hlabel{eq:125}\eeq
where
\beq
B(p^2)={\Gamma(-\epsilon/2)\over(4\pi)^{D/2}}
\left({m\over\mu}\right)^\epsilon\int_0^1dx
\left[\left(1-x(1-x){p^2\over m^2}\right)^{\epsilon/2}-1\right].
\hlabel{eq:126}\eeq
The result (\ref{eq:125}) is transverse as is required by the target-space
Lorentz invariance of (\ref{eq:124}). If we take $m$ fixed and send
$\epsilon\to0$ the function $B$ admits a finite limit meaning that the
ultraviolet divergences which are present in both diagrams
(Fig.~\ref{fig:eabosons}) cancel each other. Explicitly we get
\bns
B\big|_{\epsilon=0}&=-{1\over4\pi}\int_0^1dx\ln
\left(1-x(1-x){p^2\over m^2}\right)\\
&=-{1\over4\pi}\int_0^1dx\ln\left(-{m^2\over p^2}+x(1-x)\right)
-{1\over4\pi}\ln\left(-{p^2\over m^2}\right).\yesnumber
\hlabel{eq:127}\ens
However, this result does not admit a finite limit for $m\to0$ which signals
the presence of an infrared divergence as anticipated above. For $m\to0$ the
divergence can be directly read off from (\ref{eq:127})
\beq
\lim_{m\to0}B\big|_{\epsilon=0}\to
{1\over4\pi}\left(2-\ln\left(-{p^2\over m^2}\right)\right).
\hlabel{eq:128}\eeq
Alternatively in (\ref{eq:126}) we can first send $m\to0$ and then regularize
the infrared divergence with the dimensional regularization which is already
present
\bes
B\big|_{m=0}={1\over4\pi}\left(-{p^2\over4\pi\mu^2}\right)^{\epsilon/2}
\Gamma(-\epsilon/2)\int_0^1dx(x(1-x))^{\epsilon/2}.
\ees
Sending now $\epsilon\to0$ the infrared divergence shows up as a simple pole
in $\epsilon$
\beq
\lim_{\epsilon\to0}B\big|_{m=0}\to{1\over4\pi}
\left(2-\ln\left(-{p^2\over4\pi\mu^2}\right)-{2\over\epsilon}-\gamma\right)
\hlabel{eq:129}\eeq
where $\gamma$ is Euler's constant.

To our knowledge infrared divergences of this type have not yet
been discussed in string theory and at present we have no proof for their
cancellation. Below we will argue that these divergences are actually only
perturbative effects. Comparing (\ref{eq:128}) with (\ref{eq:129}) we can
separate out the infrared divergence and determine the finite part of $B$ to
be
\beq
B_f={1\over2\pi}.
\hlabel{eq:130}\eeq
In writing (\ref{eq:130}) we omitted the term $\ln(-p^2)$ and the other
(finite and divergent) parts which we interpret as infrared effects, for the
discussion see below.
A similar criterium for the separation of infrared divergences has been
adopted in \hcite{GRINIE} to prove the absence of a level shift in the WZWN
model at two loops. In our case (\ref{eq:130}) is actually the unique choice
which leads to a Wess-Zumino consistent anomaly as we will see in the next
section. With (\ref{eq:130}) we get for (\ref{eq:125})
\bes
\Gamma_{iab;j}{}^{cd}={\alpha\over2\pi}\delta_{[a}^d\delta_{b]}^c
\left(\eta_{ij}-{p_ip_j\over p^2}\right).
\ees
Upon adding the external legs we obtain for the boson contribution to the
effective action
\bes
\Gamma_B={\alpha\over4\pi}\int d^2\sigma\trace\left(\Omega_i
\left(\eta^{ij}-{\partial^i\partial^j\over\Box}\right)\Omega_j\right)
\ees
and by restoring the worldsheet metric we get
\beq
\Gamma_B={\alpha\over16\pi}\int d^2\sigma\gs\trace
\left[\left(D_-\Omega_+-D_+\Omega_-\right){1\over\Box_g}
\left(D_-\Omega_+-D_+\Omega_-\right)\right].
\hlabel{eq:131}\eeq
The total effective action can now be computed from
(\ref{eq:104tex}) and (\ref{eq:131}) to be
\bns
\Gamma&=\Gamma_F+\Gamma_B\\
&={\alpha\over16\pi}\int d^2\sigma\gs\trace
\left(D_+\Omega_-{1\over\Box_g}(D_+\Omega_--D_-\Omega_+)\right)\\
&={\alpha\over8\pi}\int d^2\sigma\gs\trace
\left(\Omega_-{1\over D_-}{\varepsilon^{ij}\partial_i\Omega_j\over\gs}\right)
\yesnumber\hlabel{eq:133}\ens
which is now, apart from a sign difference due to the opposite chirality of the
heterotic fermions and the $y^\alpha$, formally identical to the
effective action gotten from the integration over the heterotic fermions, see
(\ref{eq:ea100}). In particular (\ref{eq:133}) does not depend on $e_+{}^i$,
but only on the $\kappa$-invariant fields $e_-{}^i$ and $\gs$. Therefore, when
computing the $\kappa$-variation of (\ref{eq:133}) it is not necessary to vary
the world-sheet metric, but we can limit ourselves to vary the induced
connection $\Omega_i$. To understand better the non-local contributions of
this variation we vary $\Gamma_F$ and $\Gamma_B$ separately
\beb\bns
\delta_\kappa\Gamma_F&=-{\alpha\over8\pi}\int
d^2\sigma\varepsilon^{ij}\trace
\left(L\partial_i\Omega_j-R_i\Omega_j\right)\\
&-{\alpha\over8\pi}\int d^2\sigma\gs\left[\trace
\left(D_-\left[L,\Omega_+\right]{1\over\Box_g}D_-\Omega_+\right)
+\trace\left(D_-R_+{1\over\Box_g}D_-\Omega_+\right)\right]
\yesnumber\hlabel{eq:134}\\
\delta_\kappa\Gamma_B&=-{\alpha\over8\pi}\int
d^2\sigma\,2\varepsilon^{ij}\trace
\left(R_i\Omega_j\right)\\
&+{\alpha\over8\pi}\int d^2\sigma\gs\left[\trace\left(
\left(D_-\left[L,\Omega_+\right]-D_+\left[L,\Omega_-\right]\right)
{1\over\Box_g}\left(D_-\Omega_+-D_+\Omega_-\right)\right)\right.\\
&\left.+\trace\left(D_-R_+{1\over\Box_g}D_-\Omega_+\right)\right].
\yesnumber\hlabel{eq:135}
\ens\eeb
Now let us discuss the non-local terms in (\ref{eq:134}) and (\ref{eq:135});
first we notice that the non-local terms proportional to $R_+$ cancel between
(\ref{eq:134}) and (\ref{eq:135}). The term proportional to
$\left[L,\Omega_+\right]$ in (\ref{eq:134}) is cancelled by the
$\kappa$-variation of the $\Omega$-$\Omega$-$\Omega$ contribution to the
effective action gotten from the integration over the fermionic $y^\alpha$
since this term is due to the (field-dependent) Lorentz transformation
contained in the $\kappa$-transformation, and as we saw in the preceding
section (see formula (\ref{eq:118})), the $\Omega$-$\Omega$-$\Omega$
contribution does not affect the Lorentz anomaly. This is completely
analogous to the case of the heterotic fermions. The non-local contributions
in (\ref{eq:135}) which are proportional to $\left[L,\Omega_\pm\right]$ are
cancelled by the (Lorentz part of) the variation of the
$\Omega$-$\Omega$-$\Omega$ contribution to the effective action gotten by the
integration over the bosonic $y^a$, simply because the $y^a$ do not contribute
to the Lorentz anomaly. Adding up the remaining contributions, which are all
local, we get for the $\kappa$-anomaly
\bns
\Acal_L^\kappa&=-{\alpha\over8\pi}\int d^2\sigma\varepsilon^{ij}\trace
\left(L\partial_i\Omega_j+R_i\Omega_j\right)\\
&={\alpha\over16\pi}\int d^2\sigma
\varepsilon^{ij}V_i{}^AV_j{}^B\Delta^\gamma(\omega_{3L})_{\gamma BA}
\yesnumber\hlabel{eq:137}\ens
where we used the super Lorentz-Chern-Simons form defined in (\ref{eq:csdef}).

Clearly the result can also be obtained by varying directly (\ref{eq:133}) and
keeping only the local terms. The anomaly in (\ref{eq:137}) can be eliminated
in the same way as the Yang-Mills type $\kappa$-anomaly in section V. The
anomaly (\ref{eq:137}) can be cancelled if we modify once more
Eq.~(\ref{eq:Hdef}) defining a new three-form field strength $H$ according to
\beq
H=dB+{\alpha\over8\pi}\left(\omega_{3YM}-\omega_{3L}\right)
\hlabel{eq:138}\eeq
and impose on $H$ defined in (\ref{eq:138}) the constraints
\beq\bgn
H_{\alpha\beta\gamma}&=H_{ab\alpha}=0\\
H_{a\alpha\beta}&=2(\Gamma_a)_{\alpha\beta}.
\egn\hlabel{eq:139}\eeq
Notice that (\ref{eq:138}) coincides with the definition (\ref{eq:122}), i.e.
precisely the relation which ensures also the cancellation of gauge and
Lorentz anomalies.

We will comment on possible additional ``true'' one-loop $\kappa$-anomalies in
the next section. Here we would like to point out that at one-loop the
effective action can produce trivial $\kappa$-anomalies which have to be
eliminated by performing suitable local subtractions on the classical action
(\ref{eq:action}). We will illustrate this fact in the following example.

In fact, additional contributions to the one-loop effective actions can be
computed by observing that the seventh and nineth term in (\ref{eq:fullexp})
correspond formally to a shift of the connection $\Omega_{ia}{}^b$ in the
sense that they can be absorbed in the second and first term respectively by
defining formally a new Lorentz connection as
\beq
\Omegat_{ia}{}^b\equiv\Omega_{ia}{}^b-{1\over2}e_{-i}V_+{}^gT_{ga}{}^b.
\eeq
Therefore the seventh and nineth term in (\ref{eq:fullexp}) can be taken into
account by replacing in the fermionic contribution (\ref{eq:117}) and in the
bosonic contribution (\ref{eq:131}) $\Omega_i$ with $\Omegat_i$ to get
respectively $\Gammat'_F$ and $\Gammat_B$. Summing up we obtain
\beq
\Gammat'_F+\Gammat_B=\Gamma+{\alpha\over16\pi}\int d^2\sigma\gsright
\left[\left(\Omega_{+a}{}^b-2V_+{}^gT_{ga}{}^b\right)\Omegar_{-b}{}^a
+{1\over2}\delta_-^ie_{-i}V_+{}^gT_{ga}{}^bV_+{}^hT_{hb}{}^a\right],
\eeq
and the last three terms in this formula are not $\kappa$-invariant, but local.
Therefore the seventh and nineth term give rise to a trivial $\kappa$-anomaly
which has to be eliminated by redefining the classical action according to
\beq
I\to I-{\alpha\over16\pi}\int d^2\sigma\gsright
\left[\left(\Omega_{+a}{}^b-2V_+{}^gT_{ga}{}^b\right)\Omegar_{-b}{}^a
+{1\over2}\delta_-^ie_{-i}V_+{}^gT_{ga}{}^bV_+{}^hT_{hb}{}^a\right].
\hlabel{eq:129tex}\eeq
Notice that the first two cocycles in (\ref{eq:129tex}) are precisely those
which had to be subtracted in the previous section to get a diff-invariant
Lorentz {\it anomaly\/}, see (\ref{eq:cocycle2}) and (\ref{eq:108tex}); the
last cocycle in (\ref{eq:129tex}) is Lorentz invariant and is needed to cancel
a diff-anomaly from the {\it effective action\/}.

Let us now briefly comment on the infrared divergence encountered above. The
divergence is due to the presence of scalar massless bosons, the $y^a$ which
in two dimensions are known to be plagued by infrared divergences. We argue
that in the case at hand these divergences are actually perturbative effects
by reasoning as follows. In our case, in fact, the fields $y^a$ are
``essentially'' massive, in the sense that there are terms in the action
(\ref{eq:fullexp}) which are quadratic in the $y^a$
\beq
{1\over2}\int d^2\sigma\gs \,y^a\Mcal^2_{ab}(\sigma)y^b
\hlabel{eq:142}\eeq
where $\Mcal_{ab}$ is a function of the external fields. Let us assume that
there exists a configuration of the external fields such that
$\Mcal^2_{ab}(\sigma)$ becomes a constant matrix, i.e. independent of
$\sigma$, and let us also assume, for the sake of simplicity, that this matrix
is proportional to the identity
\beq
\Mcal^2_{ab}(\sigma)=\Mcal^2\delta_{ab}.
\eeq
Then, for this configuration, (\ref{eq:142}) produces a mass term for the
scalars, with mass $\Mcal$. Then no infrared regularization is required and
formula (\ref{eq:126}) becomes
\bns
B_{\epsilon=0}&=-{1\over4\pi}\int_0^1dx\ln
\left(-{\Mcal^2\over p^2}+x(1-x)\right)
-{1\over4\pi}\ln\left(-{p^2\over\Mcal^2}\right)\\
&={1\over2\pi}-{1\over4\pi}\int_0^1dx\ln
\left({1\over x(1-x)}-{p^2\over\Mcal^2}\right).
\yesnumber\hlabel{eq:143}
\ens
The integral in (\ref{eq:143}) is now convergent, but it is
{\it non-analytic\/} in the ``external fields'' $\Mcal$. The perturbative
approach we adapted to compute the Lorentz-type $\kappa$-anomaly was based on
a power series expansion in terms of polynomials in the external fields, but
clearly (\ref{eq:143}) cannot be expanded, around $\Mcal=0$, in polynomials of
$\Mcal$. If one can generalize this argument for a generic configuration of
the external fields and we guess that this is possible, then one can conclude
that an additive part of the effective action is non-analytic in the external
fields and the infrared divergences we encountered are just signals of this
non-analyticity. The $\kappa$-invariance of the non-analytic contribution to
the
effective action seems rather difficult to control, we guess that it is
actually invariant due to the fact that anomalies should always be local, and
hence analytic.

As a last remark of this section we would like to stress that extracting as
``analytic'' part from (\ref{eq:143}) the constant $1/2\pi$ turns out to
be actually the correct choice because the anomaly computed with this
constant, and only with this constant, turns out a) to be local and b) to
satisfy the Wess-Zumino consistency condition. In fact, for a different
constant the non local-terms proportional to $R_+$ would not cancel between
(\ref{eq:134}) and (\ref{eq:135}).

A cohomogical analysis of the computed $\kappa$-anomalies and a brief
discussion of the resulting SUGRA-SYM theory follows in the next section.
\vspace{1truecm}
\section{Wess-Zumino consistency condition and SUGRA-SYM theory}
As anticipated in the introduction the computation of one-loop
$\kappa$-anomalies permits, imposing their cancellation, to derive the
order-$\alpha$ corrections to the classical constraints on the superfields of
the background theory. As has been shown in \hcite{TONIN} the Wess-Zumino
consistency condition which has to be satisfied by the $\kappa$-anomalies
ensures the solvability of the Bianchi identities with these new constraints.

In this section we want to describe the main features of this method to derive
in particular the consistent order-$\alpha$ corrections to the pure
$N=1$, $D=10$ SUGRA-SYM theory and apply it to the anomalies we
have computed.

The total anomaly computed in the previous sections can be written as
\beq
\Acal_\kappa=-{\alpha\over16\pi}\int d^2\sigma
\varepsilon^{ij}V_i{}^AV_j{}^B\Delta^\gamma G_{\gamma BA}
\hlabel{eq:144}\eeq
where $G_{\gamma BA}$ are the components of the three-superform
\beq
G={1\over3!}E^AE^BE^CG_{CBA}\equiv\omega_{3YM}-\omega_{3L}
\hlabel{eq:144a}\eeq
satisfying
\beq
dG=\trace F^2-\trace R^2.
\hlabel{eq:145}\eeq
By taking for the BRS transformations of the ghosts $\kappa_{+\alpha}$
(the ghosts $\kappa_{+\alpha}$ commute between themselves,
$\kappa_{+\alpha}\kappa_{+\beta}=\kappa_{+\beta}\kappa_{+\alpha}$)
\beq
\delta_\kappa\kappa_{+\alpha}=\kappa_{+\beta}\kappa_{+\gamma}
\left(\Vsla^{\beta\varepsilon}_-\Omega_{\varepsilon\alpha}{}^\gamma
+\delta_\alpha^\beta(\Vsla\lambda)^\gamma
-\Vsla_-^{\beta\gamma}\lambda_\alpha
+4\delta_\alpha^\beta V_-{}^\gamma-
(\Gamma_g)^{\beta\gamma}(\Gamma^g)_{\alpha\varepsilon}V_-{}^\varepsilon\right),
\eeq
we can construct an on-shell nihilpotent BRS operator $\Omega_\kappa$,
satisfying
$\Omega^2_\kappa=0$ (on shell). Then the anomaly is characterized as a
(non-trivial)
cocycle of $\Omega_\kappa$ satisfying the BRS consistency condition
\beq
\Omega_\kappa\Acal_\kappa=0.
\hlabel{eq:146}\eeq
By rewriting (\ref{eq:144}) as
\bes
\Acal_\kappa=-{\alpha\over16\pi}\int d^2\sigma\varepsilon^{ij}
\partial_iZ^M\partial_jZ^N\delta_\kappa Z^LG_{LNM},
\ees
we can compute (\ref{eq:146}), which turns out to be,
modulo terms proportional to the equations of motion,
\bno
\Omega_\kappa\Acal_\kappa&=-{\alpha\over8\pi}\int d^2\sigma\varepsilon^{ij}
\partial_iZ^M\partial_jZ^N\delta_\kappa Z^L\delta_\kappa Z^P
\partial_{[P}G_{LNM)}\nonumber\\
&=-{\alpha\over32\pi}\int d^2\sigma\varepsilon^{ij}
V_i{}^AV_j{}^B\Delta^\alpha\Delta^\beta(dG)_{\beta\alpha BA}=0.
\hlabel{eq:146b}\eno

Due to the constraints (\ref{eq:Fconst}), (\ref{eq:Rconst})  with
(\ref{eq:145}) the condition (\ref{eq:146b}) reduces to
\bes
V_-{}^cV_+{}^d\Delta^\alpha\Delta^\beta
\left(\trace F^2-\trace R^2\right)_{\alpha\beta cd}=0
\ees
which, under the constraints (\ref{eq:grconst}), (\ref{eq:7}), becomes
\beq
\Delta^\alpha\Vsla_{-\alpha\gamma}
\left[\trace(\chi^\gamma\chi^\delta)-\trace(T^\gamma T^\delta)\right]
\Vsla_{+\delta\beta}\Delta^\beta=0
\hlabel{eq:151co}\eeq
where we wrote $\trace(T^\gamma T^\delta)\equiv T_{ab}{}^\gamma T^{\delta ba}$.
On-shell (\ref{eq:151co}) is identically satisfied due to Eq.~(\ref{eq:delid})
and (\ref{eq:virasoro})
so that under the constraints (\ref{eq:grconst}), (\ref{eq:7})
our anomaly satisfies the consistency condition identically.

In \hcite{TONIN} it has been shown that for a generic $G$ satisfying
(\ref{eq:146b}) the Bianchi identities can be consistently solved with the
constraints (\ref{eq:139}) and the definition $H=dB+{\alpha\over8\pi}G$.
Then the Bianchi identities (\ref{eq:123})
can be consistently solved with the constraints (\ref{eq:139}) while the
constraints (\ref{eq:6a}), (\ref{eq:Fconst}) remain unchanged. The check of
the consistency of the
Bianchi identities is straightforward, here we report the order-$\alpha$
corrected relations between the various superfields. Notice that it is not
consistent to keep $\alpha^2$-corrections in that for getting the complete
$\alpha^2$-corrections one had to compute two-loop anomalies together with
other arrangements, see the discussion in the concluding section. We get
\beb\hlabel{eq:148}\bno
T_{a\alpha}{}^\beta&={1\over4}(\Gamma_{bc})_\alpha{}^\beta T_a{}^{bc}
-{\alpha\over16\pi}(\Gamma_a)_{\alpha\varepsilon}
\left(\trace(\chi^\varepsilon\chi^\beta)
-\trace(T^\varepsilon T^\beta)\right)\\
D_\alpha T_{abc}&=(\Gamma_{[a})_{\alpha\beta}\left(-6T_{bc]}{}^\beta
-{3\alpha\over8\pi}\left(\trace(F_{bc]}\chi^\beta)-\trace(R_{bc]}T^\beta)\right)
\right)\\
D_\alpha\lambda_\beta&=-(\Gamma_g)_{\alpha\beta}D^g\phi
+\lambda_\alpha\lambda_\beta\nonumber\\
&\qquad+{1\over12}(\Gamma_{abc})_{\alpha\beta}
\left[T^{abc}+{\alpha\over64\pi}(\Gamma^{abc})_{\gamma\delta}
\left(\trace(\chi^\gamma\chi^\delta)-\trace(T^\gamma T^\delta)\right)\right]\\
R_{\alpha\beta ab}&=-{\alpha\over8\pi}(\Gamma_{[a})_{\alpha\varepsilon}
\left(\trace(\chi^\varepsilon\chi^\varphi)-
\trace(T^\varepsilon T^\varphi)\right)(\Gamma_{b]})_{\varphi\beta}\\
R_{a\alpha bc}&=2(\Gamma_a)_{\alpha\beta}T_{bc}{}^\beta
+{3\alpha\over16\pi}(\Gamma_{[a})_{\alpha\beta}
\left[\trace(F_{bc]}\chi^\beta)-\trace(R_{bc]}T^\beta)\right].
\eno\eeb
In particular we have again
\beq
H_{abc}=T_{abc}.
\hlabel{eq:148a}\eeq
With respect to the zeroth order constraints the principal feature is the
appearance of a non-vanishing $R_{\alpha\beta ab}$, which acquires now a 120
irreducible representation (irrep) of $SO(10)$, as is expected on general
grounds for non-minimal supergravity theories, see
\hcite{BBLPTJ,RARIZA,CANLEC}.
Notice that now $(\trace R^2)_{\alpha\beta\gamma\delta}$ and
$(\trace R^2)_{\alpha\beta\gamma a}$ are no longer zero, but of order $\alpha$
and hence the Wess-Zumino condition (\ref{eq:146}) is no longer satisfied
identically: it is satisfied only at first order in $\alpha$ according to our
one-loop computation. We stress again that to take $\alpha^2$-corrections into
account one had to go to two-loops.

Let us now discuss the presence of possible additional ``true'' anomalies at
first order in $\alpha$, i.e. at one loop. For this purpose it is convenient to
recall that the total $\kappa$-anomaly $\Acal_\kappa^T$ and the gauge and
Lorentz anomalies $\Acal_G$ and $\Acal_L$ satisfy on general grounds the
following coupled cohomology problem
\beb\hlabel{eq:149}\bno
\Omega_\kappa\Acal_\kappa^T=0,\qquad\Omega_L\Acal_G+\Omega_G\Acal_L&=0
\hlabel{eq:149a}\\
\Omega_G\Acal_G=0,\qquad\Omega_L\Acal_\kappa^T+\Omega_\kappa\Acal_L&=0\\
\Omega_L\Acal_L=0,\qquad\Omega_G\Acal_\kappa^T+\Omega_\kappa\Acal_G&=0
\eno\eeb
where $\Omega_\kappa$, $\Omega_G$, $\Omega_L$ are the BRS operators
associated to $\kappa$, gauge and Lorentz transformations respectively.
If we take for $\Acal^T_\kappa$ the anomaly $\Acal_\kappa$ we have found, see
Eq.~(\ref{eq:144}), and for $\Acal_G$ and $\Acal_L$ (\ref{eq:gauanom}) and
(\ref{eq:120}) respectively it is not difficult to show that all the equations
in (\ref{eq:149}) are indeed satisfied, the first equation in (\ref{eq:149a})
is nothing else than (\ref{eq:146}). Now, the gauge and Lorentz anomalies
(\ref{eq:gauanom}) and (\ref{eq:120})  are expected to be exact, i.e. not to
get higher-loop corrections and clearly they are one-loop exact, but it is not
obvious at all that Eq.~(\ref{eq:144}) presents the complete one-loop
$\kappa$-anomaly. We can in general write
\beq
\Acal_\kappa^T=\Acal_\kappa+X_\kappa
\hlabel{eq:150}\eeq
where $X_\kappa$ is a possible missing anomaly. Then (\ref{eq:150}) has to
satisfy again (\ref{eq:149}) and, using the fact that $\Acal_\kappa$ satisfies
it already, we get the conditions:
\beq\bgn
\Omega_\kappa X_\kappa&=0\\
\Omega_G X_\kappa&=0\\
\Omega_L X_\kappa&=0
\egn\hlabel{eq:151}\eeq
which means that the missing anomaly $X_\kappa$ has to be gauge and Lorentz
invariant and that it has to satisfy the $\kappa$-consistency condition
independently from $\Acal_\kappa$. Possible solutions to (\ref{eq:151}) can be
constructed as follows. We write
\beq
X_\kappa={1\over2}\int d^2\sigma\varepsilon^{ij}
V_i{}^AV_j{}^B\Delta^\gamma X_{\gamma BA}
\hlabel{eq:152}\eeq
where the two $V$'s have to be there for dimensional reasons and we take
$X_{CBA}$ to be the components of a three-superform
\bes
X={1\over3!}E^AE^BE^CX_{CBA}
\ees
which has to be {\it gauge and Lorentz invariant\/}. The $\kappa$-consistency
condition on $X_\kappa$ becomes then
\beq
\int d^2\sigma\varepsilon^{ij}V_i{}^AV_j{}^B
\Delta^\gamma\Delta^\delta(dX)_{\delta\gamma BA}=0,
\hlabel{eq:152bis}\eeq
which is equivalent to
\beb\hlabel{eq:152mul}\bno
(dX)_{\alpha\beta\gamma\delta}&=0\\
(dX)_{\alpha\beta\gamma a}&=0\\
\Delta^\beta V_-{}^a(dX)_{\alpha\beta ab}V_+{}^b\Delta^\alpha&=0.
\eno
Once the first two equations are satisfied the third one can be shown to be
equivalent to
\beq
(dX)_{\alpha\beta ab}=
(\Gamma_{[a})_{\alpha\varphi}H^{\varphi\delta}(\Gamma_{b]})_{\delta\beta}
\eeq\eeb
for some antisymmetric superfield $H^{\varphi\delta}$ belonging to the 120
dimensional irreducible representation of $SO(10)$
(see \hcite{TONIN} and the previous section). A class of solutions
of Eqs.~(\ref{eq:152mul})  can be determined as follows. Let us consider a
gauge-invariant (and Lorentz covariant) superfield $Y^{abcd}(Z)$ which is
antisymmetric in all its indices and belongs therefore to the 210 irrep of
$SO(10)$. Let us assume, moreover that the combination
$D_\alpha Y_{abcd}+2\lambda_\alpha Y_{abcd}$ does not contain the highest
1440-dimensional irrep of $SO(10)$, i.e.
\beq
\left(D_\alpha Y_{abcd}+2\lambda_\alpha Y_{abcd}\right)^{1440}=0.
\hlabel{eq:153}\eeq
Then we can construct an $X$ satisfying (\ref{eq:152bis}) in the following way:
\beq\bgn
X_{\alpha\beta\gamma}&=0\\
X_{a\alpha\beta}&=(\Gamma_{abcde})_{\alpha\beta}Y^{bcde}.
\egn\hlabel{eq:153a}\eeq
At this point it is not difficult to show that Eqs.~(\ref{eq:152mul}) determine
{\it consistently and uniquely\/} $X_{ab\alpha}$ and $X_{abc}$.

To conclude: each (gauge-invariant and Lorentz covariant) 210 irrep satisfying
(\ref{eq:153}) specifies uniquely a cocycle of the operator $\Omega_\kappa$,
and hence a possible anomaly. If $X$ can not be written as the
superdifferential of a two-superform $\Btil$, $X\not=d\Btil$, then $X$
corresponds to a non trivial cocycle, i.e. to a true anomaly (otherwise it can
be eliminated by redefining the Wess-Zumino two-form $B$). In this last
case the anomaly (\ref{eq:152}) can be eliminated by imposing on $H$, still
defined in (\ref{eq:138}), the constraints
\bns
H_{\alpha\beta\gamma}&=0\\
H_{a\alpha\beta}&=2\Gamma_{a\alpha\beta}+X_{a\alpha\beta}
\yesnumber\hlabel{eq:154}\\
H_{ab\alpha}&=X_{ab\alpha}.
\ens
In particular the relation between $H_{abc}$ and $T_{abc}$ becomes now
\beq
H_{abc}=T_{abc}+X_{abc}
\eeq
instead of (\ref{eq:148a}). Eq.~(\ref{eq:152bis}) assures again that the
Bianchi identities can be consistently solved, in particular the field
$H^{\varphi\delta}$ modifies the relations given in (\ref{eq:148}) by
additional
terms on the r.h.s., proportional to $H^{\varphi\delta}$.

Are such additional
$\kappa$-anomalies really present at one-loop in our sigma-model? The results
of Ref.~\hcite{RARIZA} could suggest that such an additional $\kappa$-anomaly
should show up. That paper deals with the solution of the Bianchi identity
\beq
dH={\alpha\over8\pi}\left(\trace F^2-\trace R^2\right)
\eeq
at second order in $\alpha^2$ (actually this paper gives a complete all order
solution of this Bianchi identity, found previously in \hcite{BBLPTO} with a
different but equivalent set of constraints for the superfields).
It turns out that an all order solution
can be obtained if one modifies the constraints on $H$ precisely according to
(\ref{eq:154}) where $X_{a\alpha\beta}$ and $X_{ab\alpha}$
are of {\it first\/} order in $\alpha$, and, in particular, at first order in
$\alpha$ the authors of \hcite{RARIZA} got for the 210 irrep $Y_{abcd}$
appearing in (\ref{eq:153a})
\beq
Y_{abcd}=c\alpha\left(R_{[abcd]}+
T_{[ab}{}^\alpha(\Gamma_{cd]})_\alpha{}^\beta\lambda_\beta\right),
\hlabel{eq:155}\eeq
where $c$ is a constant. It can easily be verified that the $Y_{abcd}$ given
in this formula verifies
indeed (\ref{eq:153}) up to order $\alpha$ and therefore the three-superform
$X$ constructed from (\ref{eq:155}) defines a cocycle of
$\Omega_\kappa$ at first order in $\alpha$. Then one could think that
in the Green-Schwarz sigma model there should actually be an additional
one-loop $\kappa$-anomaly, parametrized by (\ref{eq:155}).
However, as will be shown elsewhere \hcite{LECHNER}, the anomaly defined
uniquely through Eqs.~(\ref{eq:155}), (\ref{eq:153a})
and (\ref{eq:152mul}) is a {\it trivial\/} anomaly at
first order in $\alpha$. Correspondingly the
solution of the $H$-Bianchi identity found in \hcite{RARIZA} can be shown to
be equivalent, at first order in $\alpha$, to the solution found by us in
Eqs.~(\ref{eq:148})  and (\ref{eq:148a}) in the sense that one solution can be
mapped to the other through a redefinition of the fields of the SUGRA-SYM
theory \hcite{LECHNER}.

Therefore we expect that no non-trivial $X_\kappa$  satisfying (\ref{eq:151})
should appear at one-loop in our sigma model; correspondingly the complete
order-$\alpha$ corrections to the pure SUGRA-SYM theory are given in
Eqs.~(\ref{eq:148}), (\ref{eq:148a}) which show a complete symmetry between
the Yang-Mills and supergravity sectors. The equations of motion can be
derived in a straightforward way from those relations using standard
superspace techniques \hcite{BBLPTO}.

Clearly non-trivial anomalies satisfying (\ref{eq:151}) have to appear at
order $\alpha^2$, i.e. at two loops in the sigma model, because the Bianchi
identities with the parametrizations (\ref{eq:148}) are satisfied only at
first order in $\alpha$.
\vspace{1truecm}
\section{Conclusions}
In this paper we established firmly the presence of the super Lorentz
Chern-Simons form in the Green-Schwarz heterotic string sigma model in a
SUGRA-SYM background. It has to be present in the definition of the field
strength associated to the two-form superpotential $B$ in order to cancel a
one-loop $\kappa$-anomaly in the sigma model and also in order to cancel the
one-loop Lorentz anomaly. The absence of $\kappa$-anomalies is a consistency
requirement in the sigma model because $\kappa$-invariance ensures the
decoupling of the eight unphysical degrees of freedom of the sixteen fermionic
$\vartheta^\mu$ variables. To guarantee this decoupling also at the quantum
level we have to require the absence of $\kappa$-anomalies.

The relations (\ref{eq:122}) and (\ref{eq:123}), which entail the absence of
gauge, Lorentz and $\kappa$-anomalies at one-loop, reduce in ordinary
ten-dimensional space-time precisely to the relations which ensure the absence
of the space-time gauge and Lorentz anomalies in $N=1$, $D=10$ SUGRA-SYM
according to the Green-Schwarz mechanism \hcite{GRESCH}.

The results of Refs.~\hcite{TONIN} imply moreover that the Bianchi identity
(\ref{eq:123}) can be consistently solved with the constraints (\ref{eq:139})
and this implies in turn that one gets equations of motion in superspace which
define a supersymmetric theory.

As we observed in section VIII no other true anomalies are expected to appear
at one loop, but at two loops anomalies of the $X$-type, Eq.~(\ref{eq:151})
have to show up for the reasons explained in that section. The computation of
these two-loop anomalies would require the following technical arrangements.

\noindent a) The normal coordinate expansion, performed in section IV,
contains a chiral gauge rotation of the heterotic fermions, with parameter
$\Lambda$ given in (\ref{eq:55old}) and an (implicit) chiral Lorentz rotation
for the fermions $y^\alpha$ with parameter $\Sigma$.
These rotations, as shown in \hcite{GRIZAN} do not leave the functional
fermion integral invariant, and therefore, when making two-loop computations,
the two corresponding Wess-Zumino actions have to be taken into account.

\noindent b) {\it All\/} trivial one-loop $\kappa$-cocycles have to be
subtracted from the classical action (\ref{eq:action}) and normal coordinate
expanded up to second order in $y^A$. The trivial cocycles we found entail
a subtraction $\Delta I$ which is given by
\bns
\Delta I&=-{\alpha\over16\pi}\int d^2\sigma
\left(\gsleft\trace(A_-\Aleft_+)+\gsright\trace(\Omegar_-\Omega_+)\right)-\\
&\qquad-{\alpha\over16\pi}\int d^2\sigma\gsright V_+{}^gT_{ga}{}^b
\left(-2\Omegar_{-b}{}^a+{1\over2}\delta_-^ie_{-i}V_+{}^hT_{hb}{}^a\right).
\yesnumber\ens
Notice, however, that this does not necessarily correspond to the whole
one-loop subtraction one should make in that we did not perform a complete
one-loop analysis of the effective action.

\noindent c) The action should be normal coordinate expanded up to the
fourth-order in $y^A$. The order-$\alpha^2$ anomaly gets contributions at
one loop from the $y^2$ terms when one inserts the  new
constraints/parametrizations (\ref{eq:148}); in particular the three-form $dB$
appearing in the normal coordinate expanded action at first-order in $y^A$
has to be substituted with $H-{\alpha\over8\pi}(\omega_{3YM}-\omega_{3L})$.
The order-$\alpha^2$ anomaly gets contributions also at two loops from the $
y^3$ and $y^4$ terms in which one has to insert the old classical constraints.

It seems to us, however, that this program, even if conceptually not too
complicated, is technically rather involved.

It may also be that to make a reliable order-$\alpha^2$ computation one has to
take the conformal and $\kappa$ ghost sectors appropriately into account and
that the absence of a $D=10$ manifest Lorentz covariance can not be so easily
handled as at one loop. In particular, it may not be sufficient to impose
appropriate $SO(8)$ transversality conditions on the background fields.
With this respect the absence of a manifestly Lorentz covariant quantization
scheme constitutes a conceptual drawback.
\section*{Acknowledgements}
The authors would like to acknowledge useful discussions with A.~Bassetto,
L.~Griguolo and P.~A.~Marchetti. They would also like to thank A.~Tseytlin for
having pointed out to them Ref.~\hcite{GRINIE} and D.~Zanon for explanations
on her work. Special thanks go to M.~Grisaru for illuminating discussions.
\vskip1truecm
\appendix
\section*{Appendix A: computation of the two-gauge fields anomaly diagram}
The anomaly of the second diagram in Fig.~\ref{fig:anom}, by use of the
anomalous vertex (\ref{eq:anvertex}), is given by
\bes
\Acal_{2ij}^{HIJ}(p,q)=i\alpha\int{d^Dk\over(2\pi)^D}
\trace\left(\left(i\hat{\ksla}\gamma^3T^H\right)
{i\over\ksla}\left(\gamma_j\gammap T^J\right)
{i\over\ksla-\qsla}\left(\gamma_i\gammap T^I\right)
{i\over\ksla-\psla-\qsla}\right);
\ees
as we need to compute this integral only in the limit for $\epsilon\to0$, due
to the presence of the hatted order-$\epsilon$ $\hat{\ksla}$ we can set the
external momenta to zero to peek the $1\over\epsilon$-pole coming from
the logarithmically divergent integral over $k$.
\bns
\Acal_{2ij}^{HIJ}(p,q)&=-\alpha\int{d^Dk\over(2\pi)^D}
{k_mk_nk_rk_s\over(k^2)^3}\trace\left(T^HT^JT^I\right)
\trace\left(\gammah_m\gamma_3\gamma_n\gamma_j\gammap\gamma_r\gamma_i
\gammap\gamma_s\right)\\
&\equiv\int {d^Dk\over(k^2)^3}\Jcal^{HIJ}_{ij}(k^3,\hat{k}).
\ens
We note that the $i$, $j$ indices, being external, are implicitly barred;
moreover, the $r$ index gets barred because it is constrained by two chiral
projectors: $\gammap\gamma_r\gammam=\gammap\gammab_r\gammam$. With these
simplifications, we can rewrite the gamma-matrices trace as
\bes
\trace\left(\gammab_r\gamma_s\gamma_3\gammah_m\gamma_n
\left(\gammam\gamma_i\gamma_j\right)\right);
\ees
now we use the fact that the integral in $k$ can only produce symmetrized
contractions of $m$, $n$, $r$, $s$ indices. But since
$\eta_{\overline{\imath}\hat{\jmath}}=0$ the only possibility is
\bns
&\left(\eta_{rs}\eta_{mn}+\eta_{rn}\eta_{ms}\right)
\trace\left(\gammab_r\gamma_s\gamma_3\gammah_m\gamma_n
\left(\gammam\gamma_i\gamma_j\right)\right)\\
&=\trace\left[\left(\gammab_r\gammab^r\gamma_3\gammah_m\gammah^m
+\gammab_r\gammah_m\gamma_3\gammah^m\gammab^r\right)
\left(\gammam\gamma_i\gamma_j\right)\right]
\ens
which vanishes by using the commutation properties of the $\gamma_3$ matrix
with $\gammah_m$ and $\gammab_r$.
\vskip1truecm

\end{document}